\documentclass[prb,twocolumn,floatfix,showpacs,citeautoscript,superscriptaddress,longbibliography]{revtex4-1}

\usepackage{pdfpages}

\usepackage{amsmath}
\usepackage{amssymb}
\usepackage{amsfonts}
\usepackage{graphicx,graphics}
\usepackage{subcaption}
\usepackage{ragged2e}

\DeclareCaptionJustification{justified}{\justifying}
\newcommand{\ket}[1]{\lvert #1\rangle}
\newcommand{\bra}[1]{\langle#1 \rvert}
\newcommand{\abs}[1]{\lvert #1 \rvert}
\newcommand{\expect}[1]{\langle #1\rangle}

\begin{document}

\title{Full counting statistics and shot noise of cotunneling in quantum dots\\
  and single-molecule transistors} 
\author{Kristen Kaasbjerg}
\email{cosby@fys.ku.dk}
\affiliation{Department of Condensed Matter Physics, Weizmann Institute of
  Science, Rehovot 76100, Israel}
\author{Wolfgang Belzig}
\affiliation{Fachbereich Physik, Universit\"at Konstanz, D-78457 Konstanz, Germany}
\date{\today}

\begin{abstract}
  We develop a conceptually simple scheme based on a master-equation approach to
  evaluate the full-counting statistics (FCS) of elastic and inelastic
  off-resonant tunneling (cotunneling) in quantum dots (QDs) and molecules. We
  demonstrate the method by showing that it reproduces known results for the FCS
  and shot noise in the cotunneling regime. For a QD with an excited state, we
  obtain an analytic expression for the cumulant generating function (CGF)
  taking into account elastic and inelastic cotunneling. From the CGF we find
  that the shot noise above the inelastic threshold in the cotunneling regime is
  inherently super-Poissonian when external relaxation is weak. Furthermore, a
  complete picture of the shot noise across the different transport regimes is
  given. In the case where the excited state is a blocking state, strongly
  enhanced shot noise is predicted both in the resonant and cotunneling regimes.
\end{abstract}

\maketitle

\section{Introduction}

The full counting statistics~\cite{Lesovik:Electron} (FCS) of charge transfer in
quantum dots (QDs), nanostructures and molecules is an important component in
the characterization of the microscopic processes governing the transport. As
the FCS contains the full information about the low-frequency current
fluctuations, it provides access to all higher-order moments of the current
fluctuations and hence gives insight not encoded in the average value of the
current and the shot noise~\cite{Buttiker:ShotNoise} given by the two first
moments.

Experimentally, measurements of the FCS have been realized via real-time
detection of single-electron tunneling
events~\cite{Rimberg:RealTime,Haug:Bimodal,Gossard:Counting}, and the FCS of the
charge-transfer mechanisms in various conductors ranging from tunnel junctions
and resonant tunneling in Coulomb blockaded QDs to Andreev tunneling at a
superconductor--normal-metal interface has been
characterized~\cite{Prober:Environment,Reznikov:Measurement,Gossard:CountingStats,Hirayama:Bi,Reznikov:Detection,
  Haug:Universal,Pekola:Andreev}. Theoretical schemes to evaluate the FCS in
different transport
regimes~\cite{Lesovik:Electron,Nazarov:FCS,Reznikov:Counting,Belzig:Full}, have
successfully explained the experimentally measured FCS as well as the
observation of super-Poissonian shot noise in resonant tunneling through
QDs~\cite{Ritchie:Enhanced}. Recent theoretical and experimental work has
focused on non-Markovian effects due to coupling to external equilibrium
baths~\cite{Jauho:QuantumStochastic,Jauho:CountingStatistics} and quantum
coherent effects~\cite{Liu:NonMarkovian}, finite-frequency current
statistics~\cite{Aguado:Finite,Haug:Measurement}, the effect of the
electron-phonon interaction on the FCS in molecular
contacts~\cite{Oppen:Full,Belzig:NoiseI,Komnik:Charge,Yeyati:FullCounting,Belzig:NoiseII,Yeyati:Nonlinear,Aharony:FullCounting},
interaction
effects~\cite{Komnik:Towards,Fazio:Coulomb,Brandes:Transport,Buttiker:Factorial},
as well as the signature of Majorana bound states in the
FCS~\cite{Lutchyn:Majorana}.

Transport in the off-resonant regime where the QD levels are located outside the
bias window and separated from the chemical potentials of the leads by an energy
$\delta$, is dominated by cotunneling processes~\cite{Kouwenhoven:Electron}. In
cotunneling processes, an electron or hole tunnels, either elastically or
inelastically, through the energetically forbidden charge state and only
occupies the QD virtually. Inelastic cotunneling processes turn on at bias
voltages exceeding the energy $\Delta$ of excited QD states and leave a clear
fingerprint in the current-voltage characteristics. Such cotunneling
spectroscopy is ideal for probing excited states and their lifetime in
QDs~\cite{Gossard:CoSpectroscopy},
molecules~\cite{Zant:OPV5Excitations,Bjornholm:Electrical,Balestro:NatC60Cotunneling,Franke:Protection},
and graphene
QDs~\cite{Ensslin:Observation,Vandersypen:GateDefined,Stampfer:Probing}.
Recently, the study of energy
dissipation~\cite{Segal:Vibrational,Gefen:Statistics} and heat
transport~\cite{Schuricht:Charge} in the cotunneling regime has gained interest.
Experimentally, the shot noise in the cotunneling regime has been demonstrated
to be
super-Poissonian~\cite{Kouwenhoven:ShotNoise,Haug:NoiseEnhancement,Gossard:Noise,Umansky:Bunching,Haug:SpinDependent,Muraki:ShotNoise},
in agreement with theoretical predictions~\cite{Loss:CotunnelingNoise}. Other
theoretical studies have addressed the shot noise in the presence of cotunneling
in specific systems~\cite{Schon:Cotunneling,Krompiewski:Transport,Oreg:Enhanced}
and the signature of cotunneling-assisted sequential
tunneling~\cite{Loss:Transport,Gossard:Cotunneling} in the shot
noise~\cite{Schon:CAST}.

The evaluation of the FCS taking into account cotunneling, has been addressed
theoretically~\cite{Fazio:NonMarkovian,Schon:Intermediate,Emary:Counting}.
However, the FCS in the cotunneling regime where the interplay between elastic
and inelastic cotunneling governs the FCS, remains unexplored. Due to the
difficulty of measuring single cotunneling events without collapsing the virtual
intermediate state of a cotunneling process~\cite{Gossard:Detecting}, FCS of
cotunneling processes has not been studied experimentally. Recent theoretical
proposals for probing the cotunneling time $\tau_\text{cot} \sim
\hbar/\delta$~\cite{Gefen:Weak,Romito:Measuring}, may prove useful in that
regard.

In this paper, we develop a conceptually simple scheme based on a Markovian
master equation description to evaluate the FCS of cotunneling
processes. Compared to rigorous perturbative evaluations of shot noise and FCS
to next-to-leading order in the tunnel coupling strength $\Gamma=2\pi \rho
\abs{t}^2$~\cite{Schon:Cotunneling,Fazio:NonMarkovian,Schon:Intermediate}, the
approach outlined here does not account for renormalization and broadening of
the electronic levels due to quantum fluctuations which give rise to
non-Markovian dynamics~\cite{Fazio:NonMarkovian}. However, for $k_B T, eV \gg
\Gamma$ and in the cotunneling regime $\delta \gg \Gamma$, non-Markovian effects
are suppressed and can be safely neglected~\cite{Fazio:NonMarkovian}. Our
approach applies in these regimes, and we show that it recovers results for the
shot noise and FCS in the cotunneling regime of simple models obtained with
methods taking into account non-Markovian
effects~\cite{Schon:Cotunneling,Fazio:NonMarkovian}.

We furthermore demonstrate the method by studying the shot noise across
transport regimes in a generic model for a QD system with an excited electronic
state. In particular, we address the signature in the shot noise of
cotunneling-related transport channels as well as the impact on the shot noise
in the case where the excited state is a so-called blocking state.

\section{Master equation approach to quantum transport}

Quantum transport in QDs and molecules involving higher-order tunneling
processes between the QD and the leads, can be described with a $T$-matrix based
master equation approach~\cite{Flensberg}. In this approach, the rates for
tunneling between the QD states are evaluated using a generalized Fermi's golden
rule by expanding the $T$ matrix to a desired order in tunnel-coupling
Hamiltonian, $H_T$. To lowest and next-to-leading order, this gives rise to
\emph{sequential} and \emph{cotunneling} processes, respectively. Compared to
rigorous density-matrix approaches, where a formally exact master equation for
the reduced density matrix of the QD can be obtained and systematically expanded
in $H_T$ (see, e.g.,
Refs.~\onlinecite{Timm:Tunneling,Wegewijs:Kinetic,Wegewijs:Density}), the
$T$-matrix approach does not account for quantum effects such as broadening and
renormalization of the QD levels. It is therefore only applicable in the regime
$k_BT, eV \gg \Gamma$ as well as in the Coulomb blockade regime (cotunneling
regime), i.e. $\delta \gg \Gamma$. In these regimes, the discrepancy between the
$T$-matrix approach with proper regularized rates and exact perturbation theory
vanishes~\cite{Paaske:Broken}.

In the $T$-matrix approach, the master equation is restricted to the diagonal
elements of the reduced density matrix which are equivalent to the occupation
probabilities $p_m$ for the QD states---here labeled by a collective index
$m=(N,i)$ for charge and excited state (electronic, vibrational, ...). The
master equation governing their time evolution is given by
\begin{align}
  \label{eq:rateequation}
  \dot{p}_m & = - p_m \sum_{m'\neq m} \Gamma_{m,m'} 
                  + \sum_{m'\neq m} p_{m'} \Gamma_{m',m} .
\end{align}
Together with the normalization condition $\sum_m p_m=1$, it can be solved for
the steady-state occupation probabilities, $\dot{p}_m=0$.

Without the normalization condition, the master equation takes the form of the
matrix equation
\begin{equation}
  \label{eq:MP0}
  \dot{\mathbf{p}} = \mathbf{M} \mathbf{p} ,
\end{equation}
where the diagonal (off-diagonal) elements of $\mathbf{M}$ (an $M\times M$
matrix where $M$ is the total number of QD states) are given by the first
(second) sum in Eq.~\eqref{eq:rateequation}. The master-equation matrix
$\mathbf{M}$ is singular with the eigenvector of the zero eigenvalue
corresponding to the steady-state solution.

The transition rates $\Gamma_{mm'}$ due to tunneling are obtained from the
generalized Fermi golden rule
\begin{equation}
  \Gamma_{mm'} = \frac{2\pi}{\hbar} \sum_{i'f'} 
      \abs{ \bra{f} T \ket{i} }^2 \delta(E_f - E_i) .
\end{equation}
Here, $\ket{i/f}=\ket{m/m'}\otimes\ket{i'/f'}$ are products of QD and lead
states, the sum is over possible initial $\ket{i'}$ and final $\ket{f'}$ states
of the leads, and $T = H_T + H_T G_0 H_T + \ldots$ is the $T$-matrix with
$G_0=\tfrac{1}{E_i - H_0}$ denoting the Green function of the decoupled QD and
leads described by $H_0=H_\text{QD} + \sum_\alpha H_\alpha$, $\alpha=L,R$,
$H_\alpha = \sum_{k\sigma} \varepsilon_{\alpha k} c_{\alpha k\sigma}^\dagger
c_{\alpha k\sigma}^{\phantom\dagger}$, and $H_T = \sum_{\alpha k\sigma}
t_{\alpha k} c_{\alpha k\sigma}^\dagger d_{\alpha\sigma}^{\phantom\dagger} +
\text{h.c.}$.

The lowest-order \emph{sequential} tunneling processes connect neighboring
charge states of the QD system. In this case, the master equation takes the form
\begin{align}
  \label{eq:sequential}
  \left. \dot{p}_{N,i} \right\vert_\text{seq} & = - p_{N,i} \sum_{\alpha, j} 
      \left( \Gamma_{N+1,j \atop N,i}^{\alpha} +
        \Gamma_{N-1,j \atop N,i}^{\alpha} 
      \right)  \nonumber \\
    & \quad + \sum_{\alpha,j}  \bigg[ p_{N+1,j} \Gamma_{N,i \atop N+1,j}^{\alpha} +
        p_{N-1,j} \Gamma_{N,i \atop N-1,j}^{\alpha} \bigg].
\end{align}

In next-to-leading order, \emph{cotunneling} processes involve tunneling in and
out of two leads and may excite the QD but do not change the charge state. These
processes give rise to the following additional terms
\begin{align}
  \label{eq:cotunneling}
  \left. \dot{p}_{N,i} \right\vert_\text{cot} & = - p_{N,i} \sum_{\alpha\beta, j} 
       \Gamma_{N,ij}^{\alpha\beta} 
       + \sum_{\alpha\beta j}  p_{N,j} \Gamma_{N,ji}^{\alpha\beta} .
\end{align}
Note that the terms with $i=j$ in the two sums cancel. These terms are
associated with \emph{elastic} cotunneling processes which do not change the
state and therefore do not appear explicitly in the master equation.

In addition to transport-induced transitions, relaxation mechanisms due to
coupling to bosonic degrees of freedom of an equilibrium environment (e.g.,
phonons) give rise to additional transitions between QD states. The transition
rate for these processes is given by
\begin{equation}
  \label{eq:Gamma_relax}
  \Gamma_{mm'}^\text{rel} = \frac{\gamma_{mm'}}{\hbar} 
      \abs{ n_B(\Delta E_{mm'}) } ,
\end{equation}
where $\gamma_{mm'}$ determines the relaxation rate, $\Delta E_{mm'}=E_{m'} -
E_m$, and $n_B$ is the Bose-Einstein distribution. In the absence of
tunneling-induced transition, this results in a thermalized distribution of the
QD states.

From the steady-state solution, the current into terminal $\alpha$ can be
obtained by evaluating the net rate of electrons,
\begin{align}
  \label{eq:current}
  I_{\alpha} & = e \sum_{N, ij} p_{N, i} 
      \bigg( 
        \Gamma^{\alpha}_{N-1, j \atop N, i} -
        \Gamma^{\alpha}_{N+1, j \atop N, i}
      \bigg)  \nonumber \\
      & \quad + e \sum_{N, ij \atop \beta\neq\alpha} p_{N, i} 
      \bigg( 
        \Gamma^{\beta\alpha}_{N, ij} - \Gamma^{\alpha\beta}_{N, ij}
      \bigg)
\end{align}
where the first (second) term is the sequential (cotunneling) current.

\section{Full counting statistics}
\label{sec:FCS}

The main object of interest in counting statistics of charge transfer in QD
systems is the probability distribution $P(n,t)$ for $n$ electrons having passed
through the system from the source to the drain contact in the time interval
$t$.

In practice, it is more convenient to work with the cumulant generating function
(CGF) $\mathcal{S}(\chi, t)$ which is defined by
\begin{equation}
  \label{eq:CGF}
  e^{\mathcal{S}(\chi, t)} = \sum_n P(n, t) e^{in\chi}  ,
\end{equation}
where $\chi$ is a counting field and $\mathcal{S}(0, t) = 0$ in order to fulfill
the normalization condition $\sum_n P(n,t)=1$. From the CGF, the cumulants of
the current can be obtained as the derivatives with respect to the counting
field $\chi$, i.e. $\expect{\expect{I^m}} = \tfrac{\partial^m
  \mathcal{S}(\chi)}{\partial(i\chi)^m}\vert_{\chi\rightarrow 0}$ for the $m$'th
cumulant. The average current and the current noise are given by the first two
cumulants. The probability distribution $P(n,t)$ can be obtained by inverting
Eq.~\eqref{eq:CGF},
\begin{equation}
  \label{eq:Pn}
  P(n,t) = \frac{1}{2\pi} \int_0^{2\pi} \! d\chi \, e^{-i n\chi + \mathcal{S}(\chi,t)} ,
\end{equation}
which follows from the fact that $\mathcal{S}(\chi,t)$ is periodic in
$\chi$ with a period of $2\pi$.

In order to calculate the CGF and the full counting statistics, it is convenient
to work with the $n$-resolved probabilities, $p_m(n,t)$, for the occupation of
the states. As above, $n$ here refers to the number of electrons having
traversed the junction, and the distribution for the charge transfer is related
to the $n$-resolved probabilities as $P(n,t) = \sum_m p_m(n,t)$.

At the level of the master-equation treatment outlined in the preceding section,
the time evolution of the $n$-resolved probabilities $p_m(n, t)$ is governed by
\begin{equation}
  \label{eq:dotp_n}
  \dot{\mathbf{p}}(n,t) = \sum_{n'} \mathbf{M}(n - n') \mathbf{p}(n',t) ,
\end{equation}
where the matrix elements of $\mathbf{M}$ describe the effect of tunneling and
relaxation on the occupations $p_m(n, t)$, and it has been assumed that the
dynamics is independent on the absolute value of the counting variable $n$ and
only depends on the difference $n-n'$. For sequential and cotunneling processes,
the change in the counting variable is restricted to the values $n-n'=0,\pm 1$,
implying that $p_m(n, t)$ is only connected to the neighboring probabilities
$p_{m'}(n\pm 1, t)$ and $p_{m'\neq m} (n, t)$.

After Fourier transforming Eq.~\eqref{eq:dotp_n} to $\chi$ space, the
counting-field dependent master equation takes the form
\begin{equation}
  \label{eq:dotpchi}
  \dot{\mathbf{p}}(\chi,t) = \mathbf{M}(\chi) \mathbf{p}(\chi,t)
\end{equation}
where
\begin{equation}
  \label{eq:p_chi}
  \mathbf{p}(\chi, t) = \sum_n \mathbf{p}(n, t) e^{in\chi}  ,
\end{equation}
and similarly for $\mathbf{M}$. 

In the Markovian approximation, the CGF in the stationary limit $t\rightarrow
\infty$ can be obtained from the eigenvalue $\Lambda_\text{min}(\chi)$ of the
counting-field dependent matrix $\mathbf{M}(\chi)$ with the smallest real
part~\cite{Nazarov:FCS,Jauho:QuantumStochastic}, i.e.
\begin{equation}
  \mathcal{S}(\chi, t) = t \Lambda_\text{min}(\chi) ,
\end{equation}
where $t$ isthe measurement time. The evaluation of the FCS thus boils down to
constructing the matrix $\mathbf{M}(\chi)$ for the Markovian master
equation~\eqref{eq:dotpchi} and calculating the eigen value
$\Lambda_\text{min}(\chi)$ from which the cumulants of the current can be
obtained.

The method developed by Bagrets and Nazarov~\cite{Nazarov:FCS} applies to
sequential tunneling. In this case, the counting-field dependent master
equation~\eqref{eq:dotpchi} can be constructed by replacing the rates in the
second line of Eq.~\eqref{eq:sequential}, which reside in the off diagonal of
$\mathbf{M}$, by the counting-field dependent rates~\cite{Nazarov:FCS},
\begin{equation}
  \Gamma^\alpha_{mm'}(\chi) = \Gamma^\alpha_{mm'} e^{\pm i\chi} ,
\end{equation}
where $\pm$ is for processes into/out of the counting lead $\alpha$. 

Below we generalize this approach to cotunneling by demonstrating how to
construct the $\chi$-dependent matrix $\mathbf{M}(\chi)$ when cotunneling
processes are included. Our approach is valid in the regime where $k_B T, eV \gg
\Gamma$ or $\delta \gg \Gamma$. In other words, when the tunneling-induced
broadening $\Gamma$ is smaller than one of the other energy scales.

\subsection{$\chi$-dependent master equation for cotunneling}
\label{sec:FCS_cotunneling}

In order to derive the $\chi$-dependent master equation taking into account
cotunneling processes, we start by writing up the master equation for the
$n$-resolved probabilities,
\begin{widetext}
  \begin{align}
    \label{eq:dotp_n_cotunnel}
    \left. \dot{p}_m(n, t) \right\vert_\text{cot} & = - p_m(n, t) 
        \sum_{m', \alpha\beta} \Gamma_{mm'}^{\alpha\beta} 
        \bigg[ 1 - \delta_{\alpha\beta} \delta_{mm'} \bigg]
        \nonumber \\ & \quad
        + \sum_{m', \alpha\beta} \Gamma_{m'm}^{\alpha\beta}
        \bigg[ \delta_{\alpha\beta}(1 - \delta_{mm'})  p_{m'}(n,t)
               + \delta_{\alpha L}\delta_{\beta R} p_{m'}(n-1, t)
               + \delta_{\alpha R}\delta_{\beta L} p_{m'}(n+1, t)
        \bigg]  ,
  \end{align}
\end{widetext}
where the terms with $m=m'$ ($m\neq m'$) correspond to elastic (inelastic)
cotunneling processes and we are counting the number of electrons $n$ collected
in the left lead.

By differentiating Eq.~\eqref{eq:p_chi} with respect to time and using the
identity,
\begin{equation}
  \sum_n e^{i\chi n} p(n \pm 1, t) = e^{\mp i\chi} p(\chi, t)  ,
\end{equation}
\begin{widetext}
we find
\begin{align}
    \label{eq:dotp_chi}
    \left. \dot{p}_m(\chi, t) \right\vert_\text{cot} 
        & = - p_m(\chi, t) \sum_{m', \alpha\beta} \Gamma_{mm'}^{\alpha\beta}
        \bigg[ 1 - \delta_{\alpha\beta} \delta_{mm'} \bigg]
        + \sum_{m', \alpha\beta} \Gamma_{m'm}^{\alpha\beta} 
        \bigg[ \delta_{\alpha\beta}(1 - \delta_{mm'})  
               + \delta_{\alpha L}\delta_{\beta R} e^{i\chi}
               + \delta_{\alpha R}\delta_{\beta L} e^{-i\chi}
             \bigg] p_{m'}(\chi,t) 
        \nonumber \\ 
     & = p_m(\chi, t) \sum_{m', \alpha\beta} \Gamma_{mm'}^{\alpha\beta}
         \bigg[ \delta_{mm'} 
               \big( \delta_{\alpha\beta} + \delta_{\alpha L}\delta_{\beta R} e^{i\chi}
               + \delta_{\alpha R}\delta_{\beta L} e^{-i\chi}
                \big) - 1 \bigg] 
     \nonumber \\ & \quad 
     + \sum_{m'\neq m, \alpha\beta} \Gamma_{m'm}^{\alpha\beta} 
        \bigg[ \delta_{\alpha\beta}
               + \delta_{\alpha L}\delta_{\beta R} e^{i\chi}
               + \delta_{\alpha R}\delta_{\beta L} e^{-i\chi}
             \bigg] p_{m'}(\chi,t)  \nonumber \\
     & = p_m(\chi, t) \sum_{m', \alpha\beta} 
         \bigg[ \delta_{mm'} \Gamma_{mm}^{\alpha\beta}(\chi) 
                - \Gamma_{mm'}^{\alpha\beta} \bigg] 
     + \sum_{m'\neq m, \alpha\beta} \Gamma_{m'm}^{\alpha\beta}(\chi)
         p_{m'}(\chi,t) ,
\end{align}
where the $\chi$-dependent rates are defined by
\begin{equation}
  \label{eq:Gamma_chi}
  \Gamma_{mm'}^{\alpha\beta}(\chi) = \Gamma_{mm'}^{\alpha\beta}
      \big( \delta_{\alpha\beta} + \delta_{\alpha L}\delta_{\beta R} e^{i\chi}
            + \delta_{\alpha R}\delta_{\beta L} e^{-i\chi}   \big) .
\end{equation}
\end{widetext}
Here, the Kronecker delta $\delta_{\alpha\beta}$ in the first term ensures that
the contribution from elastic cotunneling to the rate in the second term inside
the square brackets of Eq.~\eqref{eq:dotp_chi} is canceled when
$\alpha=\beta$. These terms correspond to elastic cotunneling processes
involving only one lead and therefore do not affect the counting statistics for
the current.

The $\chi$-dependent master equation in Eq.~\eqref{eq:dotp_chi} defines the
cotunneling contribution to the matrix elements of $\mathbf{M}(\chi)$ in
Eq.~\eqref{eq:dotpchi}, and is our main formal result. It demonstrates how
cotunneling processes can be included on equal footing with sequential tunneling
processes by introducing counting-field dependent cotunneling rates
[Eq.~\eqref{eq:Gamma_chi}] in the master equation. Contrary to the conventional
master equation for cotunneling [Eq.~\eqref{eq:cotunneling}], the
$\chi$-dependent equation~\eqref{eq:dotp_chi} contains contributions from
elastic cotunneling processes [first term in the last line of
Eq.~\eqref{eq:dotp_chi}]. This is due to the fact that, while they do not change
the state of QD, they contribute to the transport and hence affect the
charge-transfer statistics.  We end by noting that the $\chi$-dependent elastic
cotunneling rates appear in the diagonal elements of $\mathbf{M}(\chi)$, while
the $\chi$-dependent inelastic cotunneling rates appear in the off diagonal.

\subsection{FCS of cotunneling}

In the following subsections, we show that our approach recovers known results
for the FCS~\cite{Fazio:NonMarkovian} and shot
noise~\cite{Loss:CotunnelingNoise,Schon:Cotunneling} of cotunneling in simple
models.

\subsubsection{Elastic cotunneling through a single level}

We start by considering the simple example of elastic cotunneling through a
single off-resonant electronic level with energy $\varepsilon_0$. In this case,
the $\chi$-dependent master equation~\eqref{eq:dotp_chi} reduces to
\begin{equation}
  \label{eq:dotp_singlelevel}
  \dot p_0(\chi,t) = p_0(\chi,t) \sum_{\alpha\beta} 
         \bigg[ \Gamma_{00}^{\alpha\beta}(\chi) - \Gamma_{00}^{\alpha\beta} \bigg] 
      \equiv M(\chi) p_0(\chi,t) ,
\end{equation}
where 
\begin{align}
  \label{eq:Gamma_elastic}
  \Gamma_{00}^{\alpha\beta} 
   = \frac{\Gamma_\alpha\Gamma_\beta}{2\pi\hbar} \int \! d\varepsilon \,  
        \frac{1}{\left( \varepsilon - \varepsilon_0 \right)^2} 
      f_{\alpha}(\varepsilon)
      \left[ 1 - f_{\beta}(\varepsilon) \right]  
\end{align}
is the rate for elastic cotunneling through the level and
$\Gamma_{00}^{\alpha\beta}(\chi)$ is the corresponding $\chi$-dependent rate
given in Eq.~\eqref{eq:Gamma_chi}. 

The $\chi$-dependent matrix $\mathbf{M}(\chi)$ is here a scalar, and the
eigenvalue $\Lambda_\text{min}(\chi)$, and hence the CGF, can be read off
directly from Eq.~\eqref{eq:dotp_singlelevel},
\begin{align}
  \mathcal{S}(\chi) & = t_0 \left[ \Gamma_{00}^{LR} (e^{i\chi} - 1) 
                 + \Gamma_{00}^{RL} (e^{-i\chi} - 1) \right] .
\end{align}
This CGF corresponds to bidirectional Poisson statistics, and is in agreement
with previous work on FCS of elastic cotunneling through a single level [see
Eq.~(7) of Ref.~\onlinecite{Fazio:NonMarkovian}].

For the current and noise we find
\begin{align}
  I & = \frac{e}{t_0} \frac{\partial \mathcal{S}}{\partial (i\chi)}\bigg\vert_{\chi=0} 
      = \Gamma_{00}^{LR} - \Gamma_{00}^{RL}   \nonumber \\
    & = \frac{\Gamma_\alpha\Gamma_\beta}{2\pi\hbar} 
        \int \! d\varepsilon \,  
          \frac{1}{\left( \varepsilon - \varepsilon_0 \right)^2} 
        \left[ f_L(\varepsilon) - f_R(\varepsilon) \right]       ,
\end{align}
and
\begin{align}
  \label{eq:S_singlelevel}
  S & = \frac{e^2}{t_0} \frac{\partial^2 
      \mathcal{S}}{\partial (i\chi)^2} \bigg\vert_{\chi=0} 
      = \Gamma_{00}^{LR} + \Gamma_{00}^{RL}   \nonumber \\
    & = \coth{ \left( \frac{eV}{2k_B T} \right)}
        \frac{\Gamma_\alpha\Gamma_\beta}{2\pi\hbar} 
        \int \! d\varepsilon \,  
          \frac{1}{\left( \varepsilon - \varepsilon_0 \right)^2} 
        \left[ f_L(\varepsilon) - f_R(\varepsilon) \right]      ,
\end{align}
respectively, where $eV = \mu_L - \mu_R$. At finite temperature, the Fano factor
$F = S/e\abs{I}$ is given by $F=\coth (eV/2k_BT)$. In the limit
$k_B T \gg eV$, the noise reduces to the equilibrium Johnson-Nyquist
noise $S = 2 G_d k_B T$ where $G_d$ is the conductance. For $eV \gg k_B T$, 
shot noise becomes dominant with a Poissonian Fano factor $F=1$ as
expected for independent tunneling processes.

\subsubsection{Elastic and inelastic cotunneling through a two-level system}
\label{sec:twolevel}

Next, we consider cotunneling trough a system which, in addition to its ground
state $\ket{0}$, has an excited electronic state $\ket{1}$ with energy $\Delta$
relative to the ground state. In this case, inelastic cotunneling processes
induce transitions between the ground and excited states. The steady-state
occupation probabilities of the states are given by (see
App.~\ref{sec:twolevel_rateequation})
\begin{equation}
  \label{eq:p_twolevel}
  p_0 = \frac{\Gamma_{10}}{\Gamma_{10} + \Gamma_{01}} \quad \text{and}
  \quad p_1 = \frac{\Gamma_{01}}{\Gamma_{10} + \Gamma_{01}} ,
\end{equation}
where $\Gamma_{ij} = \Gamma_{ij}^\text{rel} + \sum_{\alpha\beta}
\Gamma_{ij}^{\alpha\beta}$ is the total transition rate due to relaxation and
inelastic cotunneling.

To obtain the CGF, we set up the counting-field dependent master-equation matrix
following Eqs.~\eqref{eq:dotp_chi} and~\eqref{eq:Gamma_chi},
\begin{align}
  \label{eq:Mchi_twolevel}
  \mathbf{M}(\chi) & = 
   \begin{pmatrix}
     \Gamma_{00}(\chi) - \Gamma_{01} & \Gamma_{10}(\chi)  \\
     \Gamma_{01}(\chi)               & \Gamma_{11}(\chi) - \Gamma_{10}  
  \end{pmatrix} ,
\end{align}
where the counting-field dependent rates are defined as
\begin{equation}
  \Gamma_{ii}(\chi) = 
      \Gamma_{ii}^{LR} \left( e^{i\chi} - 1 \right) 
     + \Gamma_{ii}^{RL} \left( e^{-i\chi} - 1 \right) ,
\end{equation}
and
\begin{equation}
  \Gamma_{ij}(\chi) = \Gamma_{ij}^\text{rel} 
     + \Gamma^{LL}_{ij} + \Gamma^{RR}_{ij}
     + e^{i\chi} \Gamma^{LR}_{ij} 
     + e^{-i\chi} \Gamma^{RL}_{ij}  ,
\end{equation}
respectively. Note that $\Gamma_{ii}(\chi=0) = 0$, implying that the standard
master-equation matrix [Eq.~\eqref{eq:M_twolevel}] is recovered for $\chi=0$.
The derivatives of the counting-field dependent rates, which will be needed
below, are given by
\begin{align}
  \frac{\partial \Gamma(\chi)}{\partial (i\chi)}
     & = \Gamma_{LR} e^{i\chi} - \Gamma_{RL} e^{-i\chi}  \\
  \frac{\partial^2 \Gamma(\chi)}{\partial (i\chi)^2}
     & = \Gamma_{LR} e^{i\chi} + \Gamma_{RL} e^{-i\chi} ,
\end{align}
for both the elastic and inelastic cotunneling rates.

From the relevant eigenvalue of the counting-field dependent master-equation 
matrix~\eqref{eq:Mchi_twolevel}, the CGF is found to be
\begin{widetext}
\begin{align}
  \label{eq:CGF_twolevel}
  \mathcal{S} (\chi) & = \frac{t_0}{2} 
      \bigg[ 
        \Gamma_{00}(\chi) + \Gamma_{11}(\chi) - \Gamma_{10} - \Gamma_{01}
     + \sqrt{\left[ \Gamma_{01} - \Gamma_{10} - \Gamma_{00}(\chi) + \Gamma_{11}(\chi) \right]^2
            + 4 \Gamma_{01}(\chi) \Gamma_{10}(\chi) }
      \bigg]  .
\end{align}
\end{widetext}
This CFG is a new result and describes the FCS of combined elastic and inelastic
cotunneling. The FCS can be interpreted as arising due to switching between
different bidirectional Poisson statistics, as discussed further below.

The current and shot noise can be obtained from the $\chi$-derivatives of the
CGF. In agreement with the standard master-equation calculation in
App.~\ref{sec:twolevel_rateequation}, we find for the current
\begin{align}
  \label{eq:I_twolevel}
  I & =  \frac{e}{t_0} \frac{\partial \mathcal{S}}{\partial (i\chi)}\bigg\vert_{\chi=0} 
        \nonumber \\
  & = p_0 \partial_{i\chi} \Gamma_{00}
      + p_1 \partial_{i\chi} \Gamma_{11}
      + p_0 \partial_{i\chi} \Gamma_{01}
      + p_1 \partial_{i\chi} \Gamma_{10} 
    \nonumber \\
  & = I_\text{el} + I_\text{inel} , 
\end{align}
where $I_\text{el}$ and $I_\text{inel}$ are elastic and inelastic contributions
given by the two first and two last terms in the second line, respectively. 

The noise given by the second derivative of the CGF is found to be
\begin{align}
  \label{eq:S_twolevel}
  S & = \frac{e^2}{t_0} \frac{\partial^2 \mathcal{S}}{\partial (i\chi)^2} \bigg\vert_{\chi=0} 
  \nonumber \\
  & = p_0 ( \partial_{i\chi}^2 \Gamma_{00} + \partial_{i\chi}^2 \Gamma_{01}) 
      + p_1 ( \partial_{i\chi}^2 \Gamma_{11} + \partial_{i\chi}^2 \Gamma_{10}) 
      \nonumber \\
  & \quad + \frac{2}{\Gamma_{01} + \Gamma_{10}} 
        \bigg[ \partial_{i\chi} \Gamma_{01} \partial_{i\chi} \Gamma_{10}
          - \partial_{i\chi} \Gamma_{00}\partial_{i\chi} \Gamma_{11}
        \bigg.    \nonumber \\
  & \quad \bigg.  
      + \left( \partial_{i\chi} \Gamma_{00} + \partial_{i\chi} \Gamma_{11} \right) I
      - I^2  \bigg]
  \nonumber \\
  & = S_\text{Poisson} + \Delta S ,
\end{align}
where the two terms in the second line describe equilibrium Johnson-Nyquist and
Poissonian shot noise and the term in the square brackets is responsible
for a non-Poissonian correction $\Delta S$ at bias voltages larger than the
inelastic threshold $V>\Delta$. We note that this result for the shot noise is
in agreement with Ref.~\onlinecite{Loss:CotunnelingNoise} [their Eq.~(5.5)].

Due to the factor in front of the square brackets, the non-Poissonian correction
$\Delta S$ diverges for $\Gamma_{01},\Gamma_{10} \rightarrow 0$ if, at the same
time, $\Gamma_{00}\neq\Gamma_{11}$. The diverging super-Poissonian noise can be
understood as follows. In the limit $\Gamma_{01},\Gamma_{10} \rightarrow 0$,
i.e. for negligible environmental relaxation and vanishing inelastic cotunneling
rates, inelastic cotunneling processes change the state of the system at a rate
that is slow compared to the rate of elastic cotunneling processes. As the
latter dominates the current, this results in a current that switches between
different values when $\Gamma_{00}\neq\Gamma_{11}$. Such telegraphic switching
between two transport channels with different conductance naturally produces
super-Poissonian shot noise.

\subsubsection{Quantum dot with a spin-split level}

The shot noise in a QD with a spin-split level has previously been studied with
the real-time diagrammatic method in Ref.~\onlinecite{Schon:Cotunneling}. This
goes beyond the $T$-matrix approach adopted here by accounting for the
broadening and renormalization of the QD levels. Here, we show that the two
methods yield identical results for the noise in the regime $k_BT\gg\Gamma$. For
$k_B T \sim \Gamma$, the disagreement between the two methods amounts to a
small quantitative difference.

The Hamiltonian of the QD is given by
\begin{equation}
  H_\text{QD} = \sum_\sigma \varepsilon_\sigma 
                c_\sigma^\dagger c_\sigma^{\phantom\dagger}
                + U n_\uparrow n_\downarrow ,
\end{equation}
where $\varepsilon_\sigma$ is the spin-dependent level position and $U$ is the
Coulomb interaction for double occupancy of the QD. The sequential and
cotunneling rates are calculated as outlined in detail in
Sec.~\ref{sec:QD_rates} below.

In Fig.~\ref{fig:schon} we show the Fano factor obtained with our approach as a
function of bias voltage and for different values of the ratio $\Gamma / k_B T$
at fixed temperature, $k_BT=0.1$. With the parameters specified in the caption
of Fig.~\ref{fig:schon}, the spin-down level is filled at $V=0$. The inelastic
spin-flip channel opens at $eV = \Delta = \varepsilon_\uparrow -
\varepsilon_\downarrow$. At $eV/2= \varepsilon_\downarrow$ ($eV/2 =
\varepsilon_\uparrow + U$), the removal (addition) energy for the spin-down
(spin-up) level becomes resonant with the chemical potentials of the drain
(source) electrode.
\begin{figure}[!t]
  \includegraphics[width=0.6\linewidth]{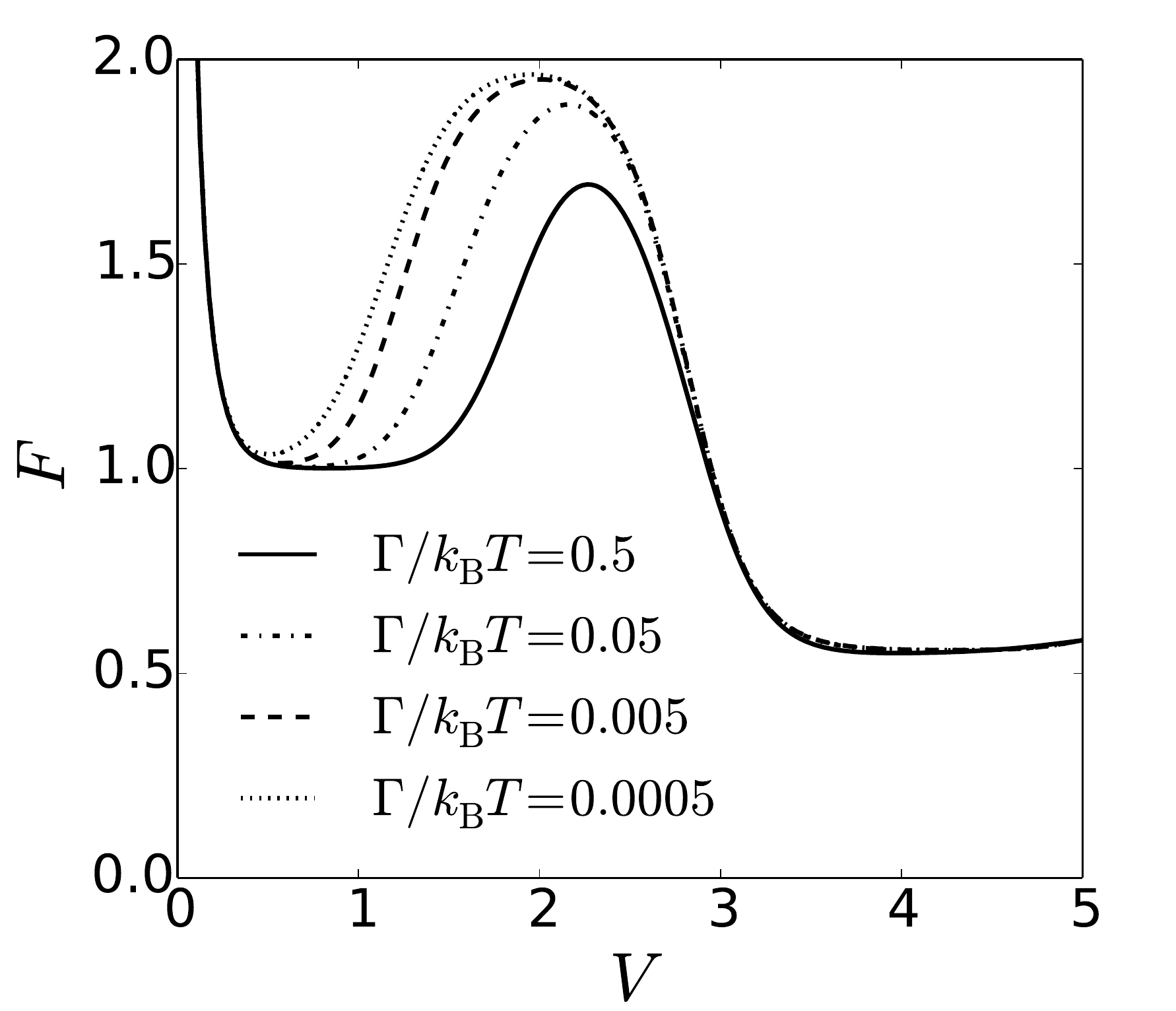}
  \caption{Fano factor for a QD with a spin-split level for different values of
    the ratio $\Gamma/k_BT$ near the cross over between the cotunneling and
    resonance regimes (see Fig.~3 in Ref.~\onlinecite{Schon:Cotunneling} for
    comparison). Parameters: $\varepsilon_\downarrow=-1.5$,
    $\varepsilon_\uparrow=0.5$, $U=4$, $k_BT=0.1$, $\Gamma_{L/R}=\Gamma / 2$,
    $\mu_{L/R} = \pm eV/2$.}
\label{fig:schon}
\end{figure}

As mentioned above, the agreement with the real-time diagrammatic approach of
Ref.~\onlinecite{Schon:Cotunneling} is perfect for $k_BT \gg \Gamma$. Remarkably
the agreement is even quite good for $k_BT \gtrsim \Gamma$ (for the present
parameters less than 10\% in the whole bias range for $\Gamma/k_BT=0.5$). For
small values of $\Gamma$, the current and noise are dominated by thermally
activated sequential tunneling through the two spin levels giving rise to
super-Poissonian noise~\cite{Belzig:Full}. At larger values of $\Gamma$, the
change in the Fano factor is due to cotunneling processes. At low bias, elastic
cotunneling dominates and results in $F=1$. Above the inelastic cotunneling
threshold at $eV=\Delta$, the state with an electron in the spin-up level
becomes populated. After a cotunneling-induced spin flip
($\downarrow \rightarrow \uparrow$), the fact that the spin-up level is located
inside the bias window, $\varepsilon_\uparrow < eV$, opens for transport through
the QD via sequential tunneling. This transport channel remains open until the
QD relaxes to the spin-down state via a sequential tunneling or an inelastic
cotunneling processes. The transport thus switches between being dominated by
elastic cotunneling through the $\downarrow$ level and sequential tunneling
through the $\uparrow$ level. As a consequence of this alternating change in the
transport mechanism, the noise becomes super Poissonian.

\section{Quantum dots and molecules with excited electronic states}

In this section, we consider the shot noise of a QD with an excited electronic
state. This could, for example, be a low-energy QD level, a molecular orbital or
spin excitations (spin manifolds). We here give a complete picture of the shot
noise across the different transport regimes and study, e.g., the signature of
the interplay between cotunneling and cotunneling-assisted sequential tunneling
(COSET)~\cite{Loss:Transport,Gossard:Cotunneling} in the shot noise. We
furthermore consider the situation where the excited state is a so-called
blocking state. This situation is familiar from, e.g., quantum dots and
molecules with excited states or broken
degeneracies~\cite{Leijnse:Asymmetric,Paaske:Broken,Kaasbjerg:ImageCharge}. In
the resonance regime outside the Coulomb blockaded regions, blocking states give
rise to pronounced negative differential conductance (NDC) and strong
super-Poissonian noise~\cite{Schon:Super}. In the cotunneling and COSET regimes,
the effect of the blocking state on the shot noise has, so far, not been
studied, and we find that the noise changes qualitatively in the presence of the
blocking state.

\subsection{Generic model and transition rates}
\label{sec:QD_rates}

We consider a spinless model for a QD where the $N$ electron configuration has
an excited electronic state. The states of the QD are described by a set of
generic many-body states
\begin{equation}
   \ket{N-1},\; \ket{Na},\; \ket{Nb}, \;\ket{N+1} ,
\end{equation}
where $N$ refers to the number of electrons on the QD. In a microscopic
description of the QD, the states and their energies result from the
diagonalization of the underlying microscopic Hamiltonian. The latter are here
parametrized as
\begin{align}
  E_{N-1} & = 0 , \quad E_{Na(b)} = \tilde{\varepsilon}_0\, (+\Delta) \nonumber \\
  \quad \text{and} & \quad E_{N+1} = 2 \tilde{\varepsilon}_0 + U    .
\end{align}
where $\Delta$ is the energy of the excited $N$-particle state $\ket{Nb}$
relative to the ground state $\ket{Na}$ with energy $\tilde{\varepsilon}_0 =
\varepsilon_0 - e V_\text{g}$ (relative to the $N-1$ state), and $U$ is Coulomb
energy associated with the addition of an electron to the $N$-particle
state. With this parametrization, the addition and removal energies of the
$N$-particle ground state become $E_{N+1} - E_{Na} = \tilde{\varepsilon}_0 + U$
and $E_{Na} - E_{N-1} = \tilde{\varepsilon}_0$, respectively, implying that the
transport gap of the QD is given by $U$.

To take into account sequential and cotunneling processes, we expand the
$T$ matrix to second order in $H_T$ in the calculation of the transition rates.

The sequential tunneling rates for adding and removing an electron from the QD
are given by 
\begin{align}
  \Gamma_{N-1,Na/b}^\alpha
   & = \frac{\Gamma_\alpha}{\hbar} \abs{M_{N-1,Na/b}^\alpha}^2 f_\alpha(\varepsilon_{a/b}) \\
  \Gamma_{Na/b, N+1}^\alpha 
   & = \frac{\Gamma_\alpha}{\hbar} \abs{M_{Na/b,N+1}^\alpha}^2 f_\alpha(\varepsilon_{a/b} + U)
\end{align}
and
\begin{align}
  \Gamma_{Na/b,N-1}^\alpha
   & = \frac{\Gamma_\alpha}{\hbar} \abs{M_{N-1,Na/b}^\alpha}^2 \left[1 - f_\alpha(\varepsilon_{a/b}) \right] \\
  \Gamma_{N+1, Na/b}^\alpha 
   & = \frac{\Gamma_\alpha}{\hbar} \abs{M_{Na/b,N+1}^\alpha}^2 \left[1 - f_\alpha(\varepsilon_{a/b} + U) \right] ,
\end{align}
respectively, where $\Gamma_\alpha = 2\pi\rho_\alpha\abs{t_\alpha}^2$ is the
tunnel broadening, $\varepsilon_{a(b)} = \tilde{\varepsilon}_0\, (+ \Delta)$,
and the matrix elements between the many-body states are given by
\begin{align}
  \label{eq:M}
  M_{Ni,N-1}^{\alpha} & = \bra{N-1} d_\alpha^{\phantom\dagger} \ket{N i}
           = M_{N-1,Ni}^{\alpha*} \\
  M_{Ni,N+1}^\alpha & = \bra{N+1} d_\alpha^\dagger \ket{N i}
           = M_{N+1,Ni}^{\alpha*} .
\end{align}
Here, $d_\alpha^\dagger, d_\alpha^{\phantom\dagger}$ denote the creation and
annihilation operators for the single-particle states in the QD system that
couple to lead $\alpha$. For a given microscopic model, the matrix elements can
be obtained from the many-body states. Here we treat them as tuneable
parameters.

The elastic cotunneling rates for the different states are given by
\begin{widetext}
\begin{align}
  \Gamma_{N-1}^{\alpha\beta}
   & = \frac{\Gamma_\alpha \Gamma_\beta}{2\pi\hbar} 
       \int \! d\varepsilon \,  
       \left\vert
         \frac{M_{N-1,Na}^\beta M_{Na,N-1}^\alpha}{\varepsilon - \tilde{\varepsilon}_0} +
         \frac{M_{N-1,Nb}^\beta M_{Nb,N-1}^\alpha}{\varepsilon - \tilde{\varepsilon}_0 - \Delta}
       \right\vert^2 
       f_{\alpha}(\varepsilon) \left[ 1 - f_{\beta}(\varepsilon) \right] 
  \label{eq:Gamma_elastic1} \\
  \Gamma_{N,a}^{\alpha\beta}
   & = \frac{\Gamma_\alpha \Gamma_\beta}{2\pi\hbar} 
       \int \! d\varepsilon \,  
       \left\vert
         \frac{M_{Na,N-1}^\alpha M_{N-1,Na}^\beta}{\varepsilon - \tilde{\varepsilon}_0} -
         \frac{M_{Na,N+1}^\beta M_{N+1,Na}^\alpha}{\varepsilon - \tilde{\varepsilon}_0 - U}
       \right\vert^2 
       f_{\alpha}(\varepsilon) \left[ 1 - f_{\beta}(\varepsilon) \right] 
  \label{eq:Gamma_elastic2} \\
  \Gamma_{N,b}^{\alpha\beta}
   & = \frac{\Gamma_\alpha \Gamma_\beta}{2\pi\hbar} 
       \int \! d\varepsilon \,  
       \left\vert
         \frac{M_{Nb,N-1}^\alpha M_{N-1,Nb}^\beta}{\varepsilon - \tilde{\varepsilon}_0 - \Delta} -
         \frac{M_{Nb,N+1}^\beta M_{N+1,Nb}^\alpha}{\varepsilon - \tilde{\varepsilon}_0 - U + \Delta} 
       \right\vert^2 
       f_{\alpha}(\varepsilon) \left[ 1 - f_{\beta}(\varepsilon) \right] 
  \label{eq:Gamma_elastic3} \\
  \Gamma_{N+1}^{\alpha\beta}
   & = \frac{\Gamma_\alpha \Gamma_\beta}{2\pi\hbar} 
       \int \! d\varepsilon \,  
       \left\vert
         \frac{M_{N+1,Na}^\alpha M_{Na,N+1}^\beta}{\varepsilon - \tilde{\varepsilon}_0 - U} +
         \frac{M_{N+1,Nb}^\alpha M_{Nb,N+1}^\beta}{\varepsilon - \tilde{\varepsilon}_0 - U + \Delta}
       \right\vert^2 
       f_{\alpha}(\varepsilon) \left[ 1 - f_{\beta}(\varepsilon) \right] .
  \label{eq:Gamma_elastic4}
\end{align}
The inelastic cotunneling rates between the ground and excited state of the
$N$-electron configuration are given by
\begin{align}
  \label{eq:Gamma_inelastic1}
  \Gamma_{N,ab}^{\alpha\beta}
   & = \frac{\Gamma_a\Gamma_b}{2\pi\hbar} 
       \int \! d\varepsilon \,  
       \left\vert
         \frac{M_{Nb,N-1}^\alpha M_{N-1,Na}^\beta}{ \varepsilon - \tilde{\varepsilon}_0 - \Delta} -
         \frac{M_{Nb,N+1}^\beta M_{N+1,Na}^\alpha}{\varepsilon - \tilde{\varepsilon}_0 - U}
       \right\vert^2 
       f_{\alpha}(\varepsilon) \left[ 1 - f_{\beta}(\varepsilon - \Delta) \right] \\
  \label{eq:Gamma_inelastic2}
  \Gamma_{N,ba}^{\alpha\beta}
   & = \frac{\Gamma_a\Gamma_b}{2\pi\hbar} 
       \int \! d\varepsilon \,  
       \left\vert
         \frac{M_{Na,N-1}^\alpha M_{N-1,Nb}^\beta}{\varepsilon - \tilde{\varepsilon}_0} -
         \frac{M_{Na,N+1}^\alpha M_{N+1,Nb}^\alpha}{\varepsilon - \tilde{\varepsilon}_0 - U + \Delta}
       \right\vert^2 
       f_{\alpha}(\varepsilon) \left[ 1 - f_{\beta}(\varepsilon + \Delta)
       \right] .
\end{align}
\end{widetext}
\begin{figure*}[!t]
\captionsetup[sub]{justification=centering,position=top,font=small,margin=0pt,skip=-1pt}
  \subcaptionbox*{\bf(a)}[0.25\linewidth]{
    \includegraphics[width=0.25\linewidth]{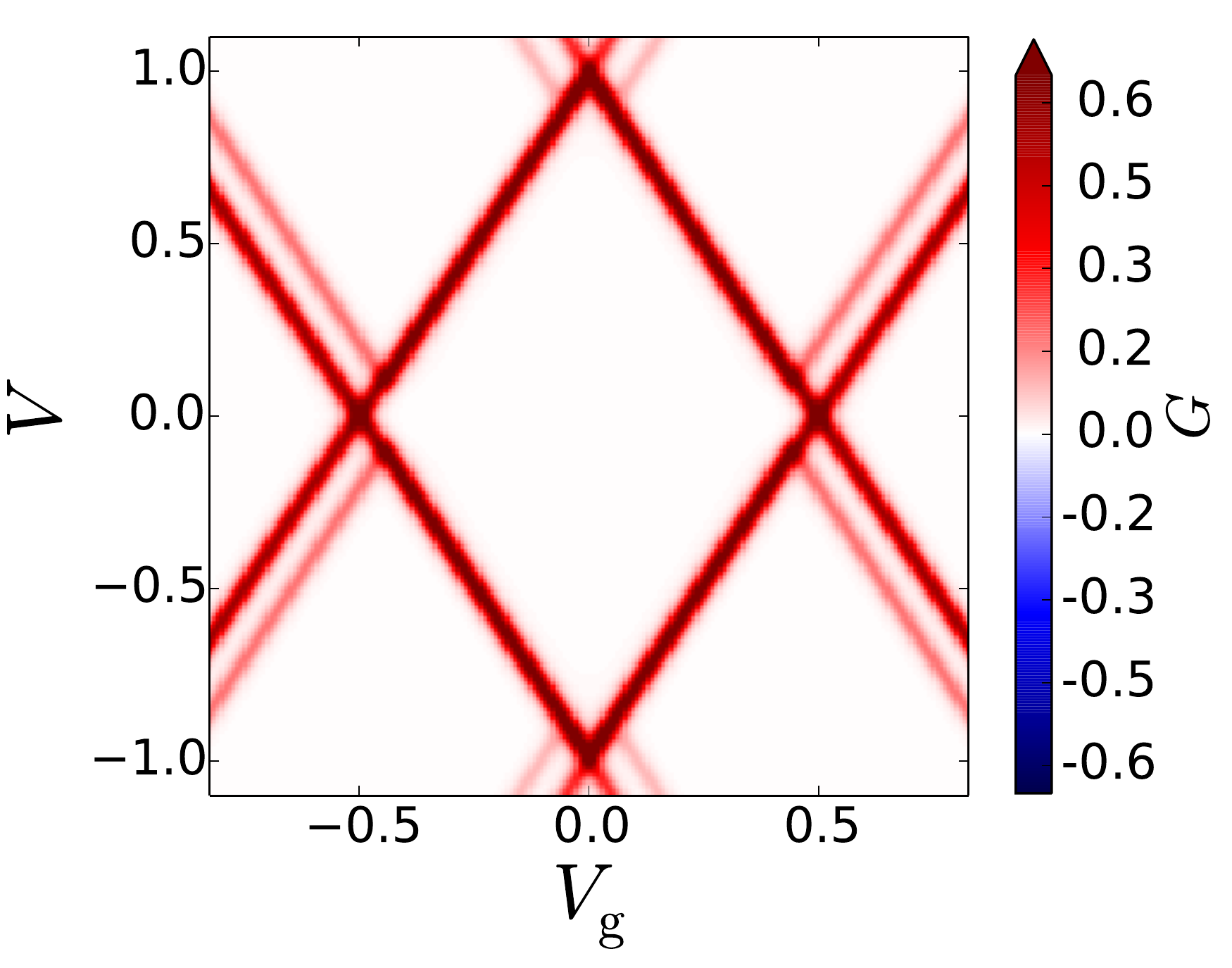}}
    \hspace{1cm}
  \subcaptionbox*{\bf(c)}[0.25\linewidth]{
    \includegraphics[width=0.25\linewidth]{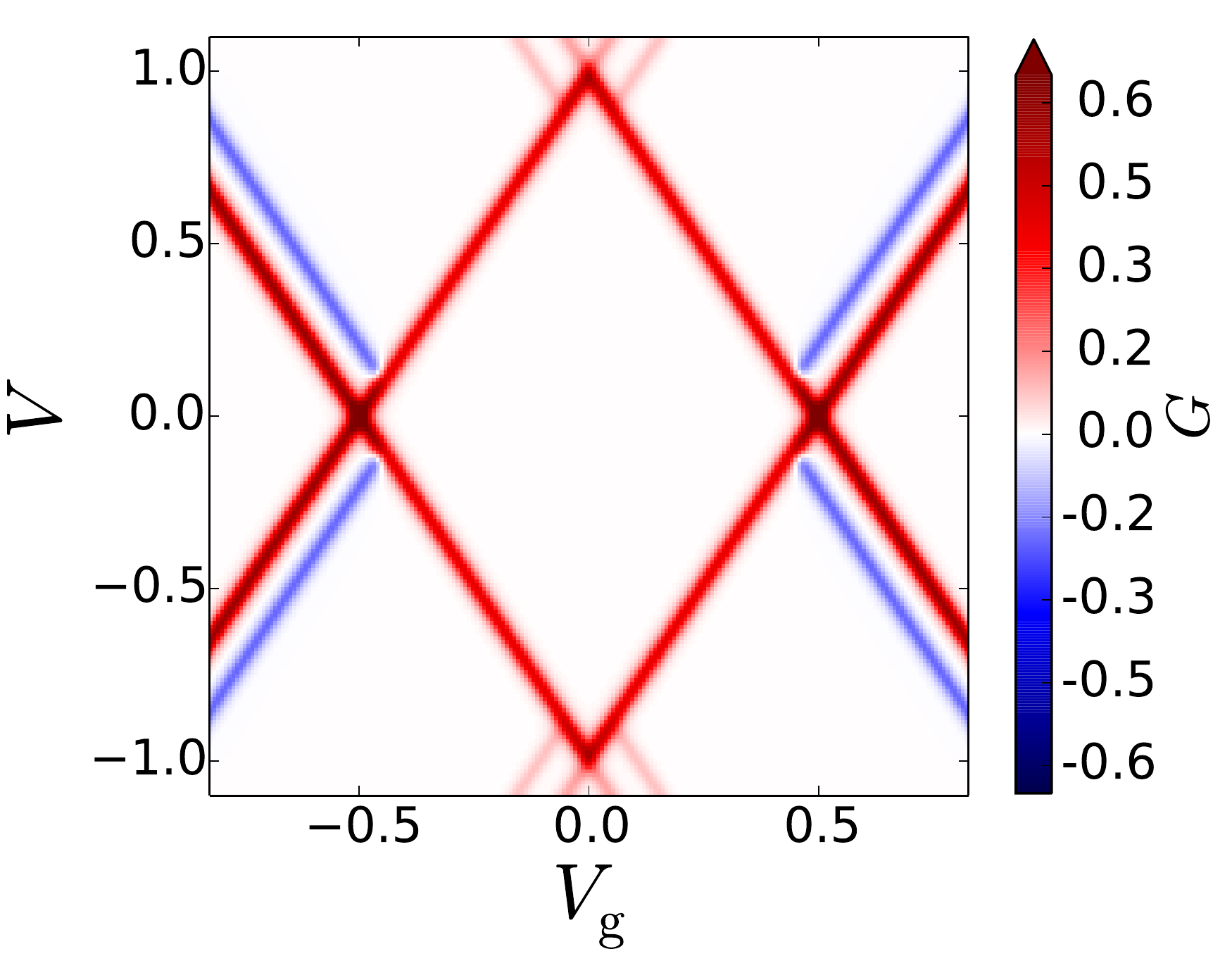}}
    \hspace{1cm}
  \subcaptionbox*{\bf(e)}[0.25\linewidth]{
    \includegraphics[width=0.25\linewidth]{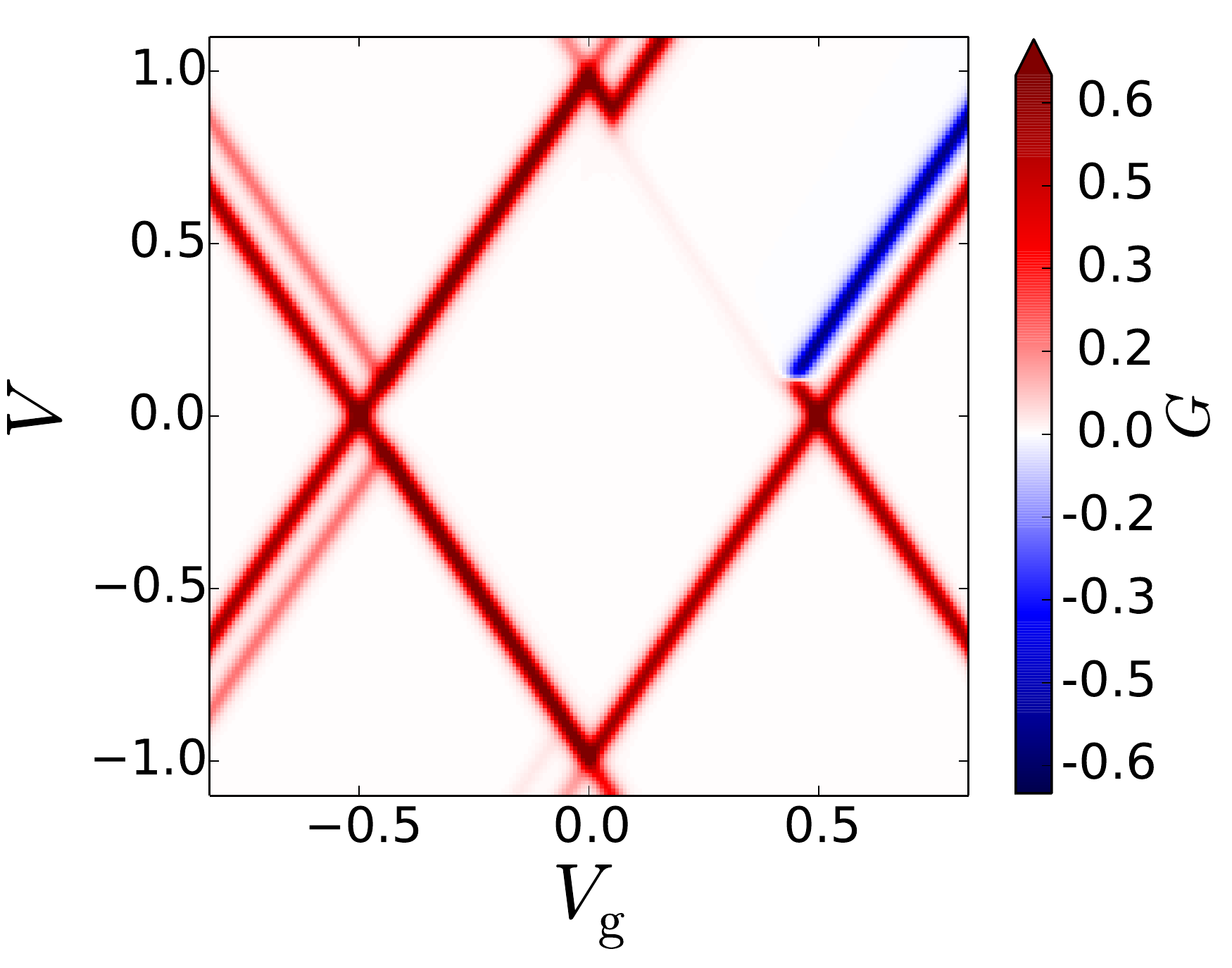}}
    \hspace{3cm}
  \subcaptionbox*{\bf(b)}[0.25\linewidth]{
    \includegraphics[width=0.25\linewidth]{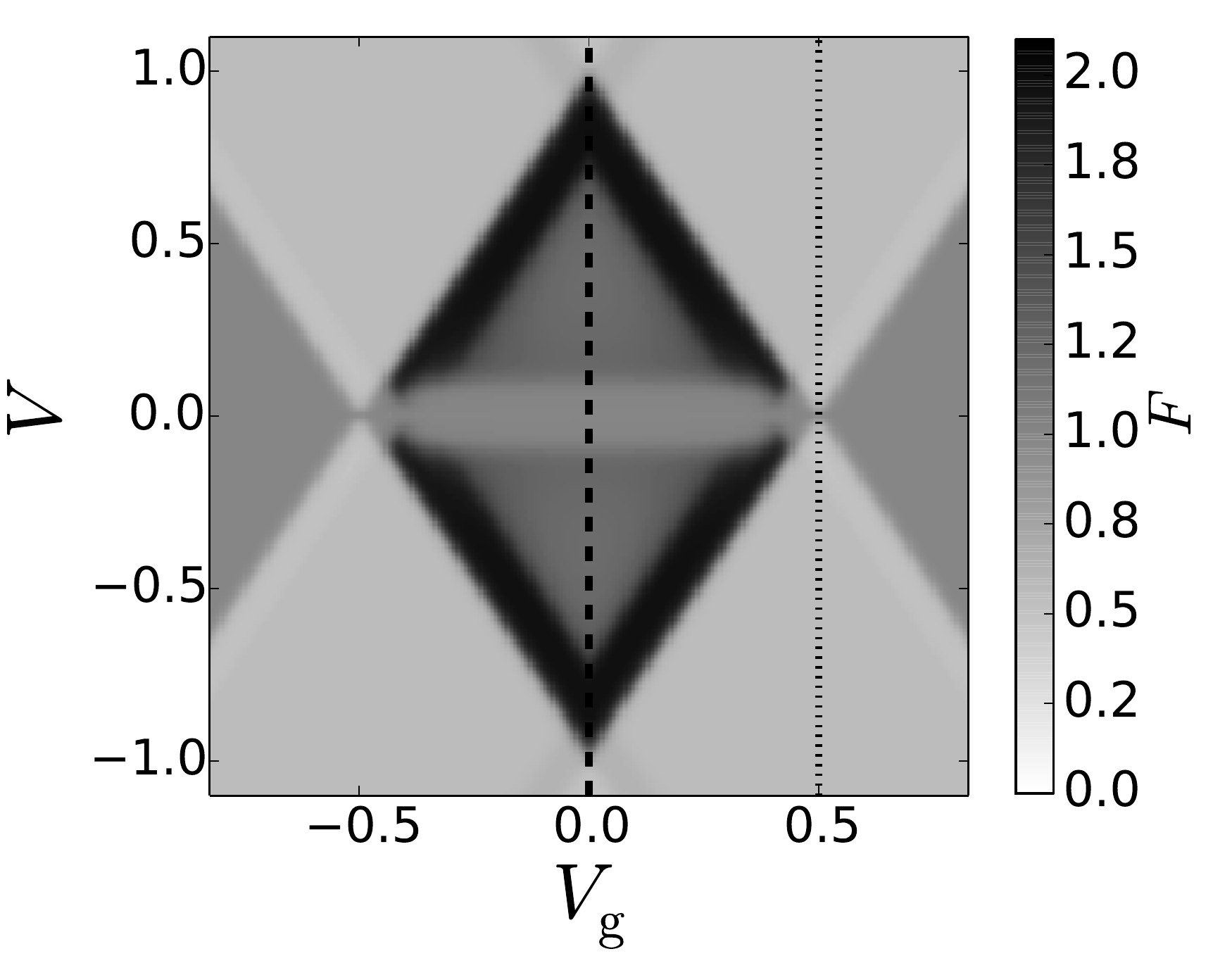}}
    \hspace{1cm}
  \subcaptionbox*{\bf(d)}[0.25\linewidth]{
    \includegraphics[width=0.25\linewidth]{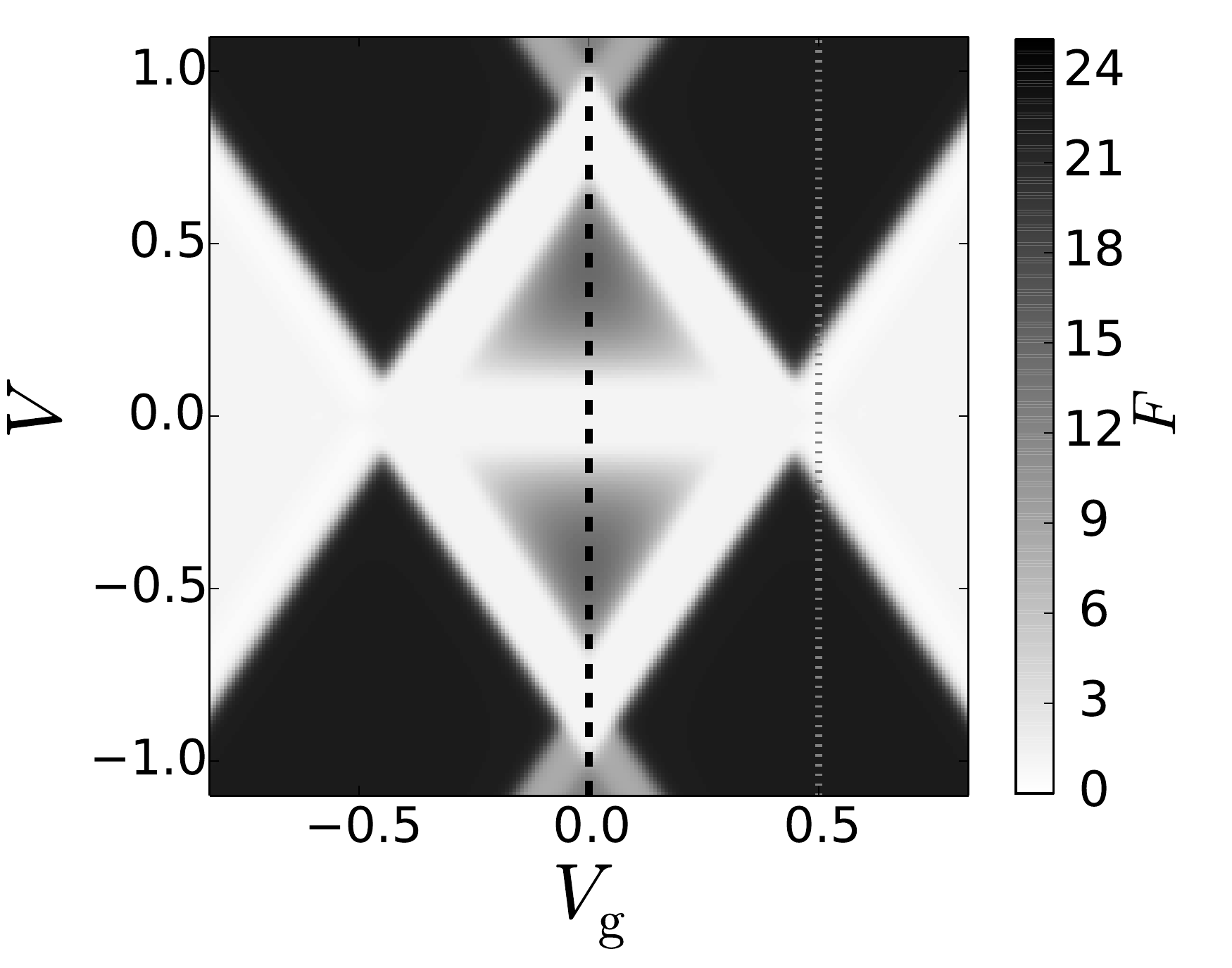}}
    \hspace{1cm}
  \subcaptionbox*{\bf(f)}[0.25\linewidth]{
    \includegraphics[width=0.25\linewidth]{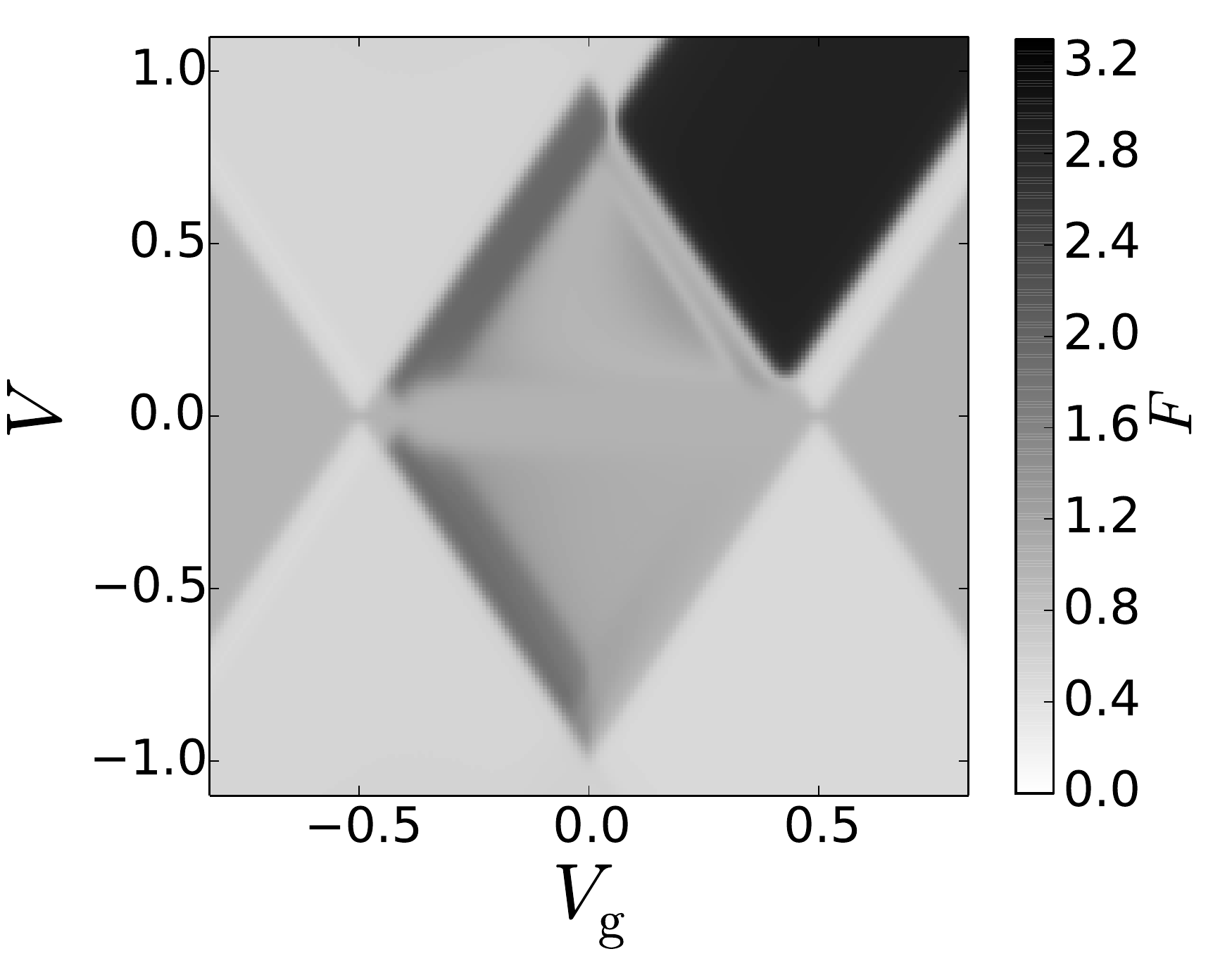}}
  \caption{(Color online) Stability diagrams showing the differential
    conductance $G=dI/dV$ (upper row) and the Fano factor $F=S/e\abs{I}$ with
    the thermal contribution to the noise subtracted (lower row) as a function
    of gate and source-drain voltage for different situations. Note the
    different color scales in the lower row. (a),(b) The ground and excited
    $N$-particle states have identical matrix elements to the $N\pm 1$ states,
    $M^{L/R}_{Na/b,N\pm1}=1$. (c),(d) and (e),(f) The excited state is a
    blocking state with small matrix elements to the $N\pm1$ states: (c),(d)
    $M^{L/R}_{N\pm 1,Nb} = 0.1$, and (e),(f) $M^{L}_{Nb,N+1}=0.1$. Matrix
    elements not specified are equal to unity. Parameters (in units of $U$):
    $\varepsilon_0=-1/2$, $U=1$, $\Delta=0.1$, $\Gamma_{L/R}=0.001$,
    $k_BT=0.01$, $\mu_{L/R} = \pm V/2$.}
\label{fig:twolevel_stability}
\end{figure*}
We evaluate the cotunneling rates at finite temperature and bias with the
commonly applied regularization scheme described in
App.~\ref{sec:regularization}.

A situation which resembles the conditions for strong super-Poissonian noise
discussed below Eq.~\eqref{eq:S_twolevel}, is realized if the excited state
$\ket{Nb}$ is a blocking state which is characterized by having small matrix
elements to the other states, i.e. $M_{Nb,N\pm1}^\alpha \ll 1$. This leads to
strongly reduced cotunneling rates and implies that the inelastic rates and the
elastic rate for $\ket{Nb}$ are reduced compared to the elastic cotunneling rate
for state $\ket{Na}$. Strong super-Poissonian noise is therefore expected. Since
the matrix elements to both the $N\pm1$ states have to be small in order to
suppress the elastic cotunneling rate for $\ket{Nb}$, the shot noise is highly
sensitive to the properties of the blocking state via its matrix elements with
the neighboring charge states.

\subsection{Stability diagrams and shot noise}

We now study the current, conductance, shot noise, and Fano factor for different
situations for the blocking property of the excited state $\ket{Nb}$. In the
upper row of Fig.~\ref{fig:twolevel_stability} we show the differential
conductance as a function of gate and source-drain bias voltage---also referred
to as charge-stability diagrams---for the cases without
[\ref{fig:twolevel_stability}(a)] and with [\ref{fig:twolevel_stability}(c) and
\ref{fig:twolevel_stability}(e)] a blocking state. For the latter, the two plots
in Figs.~\ref{fig:twolevel_stability}(c) and \ref{fig:twolevel_stability}(e)
correspond to different situations for the matrix elements involving $\ket{Nb}$
(see caption of Fig.~\ref{fig:twolevel_stability} for details). Inside the
Coulomb blockaded regions where sequential tunneling is suppressed, the current
is dominated by cotunneling processes. Due to the linear color scale in the
figures, the cotunneling features in the conductance are not visible.

Outside the blockaded regions where sequential tunneling dominate the current,
excitation lines going out from the central blockaded region appear at voltages
where the excited state enters the bias window. Some of these lines show
pronounced NDC when the excited state is a blocking state. Depending on details
of the matrix elements for the blocking state, NDC occurs either for both signs
[\ref{fig:twolevel_stability}(c)] or one sign [\ref{fig:twolevel_stability}(e)]
of the gate and bias voltages, and may completely suppress the current
[\ref{fig:twolevel_stability}(e)].
\begin{figure*}[!t]
\captionsetup[sub]{justification=centering,position=top,font=small,margin=0pt,skip=-2pt}
  \subcaptionbox*{\hspace{.3cm}\bf(a)}[0.25\linewidth]{
    \includegraphics[width=0.25\linewidth]{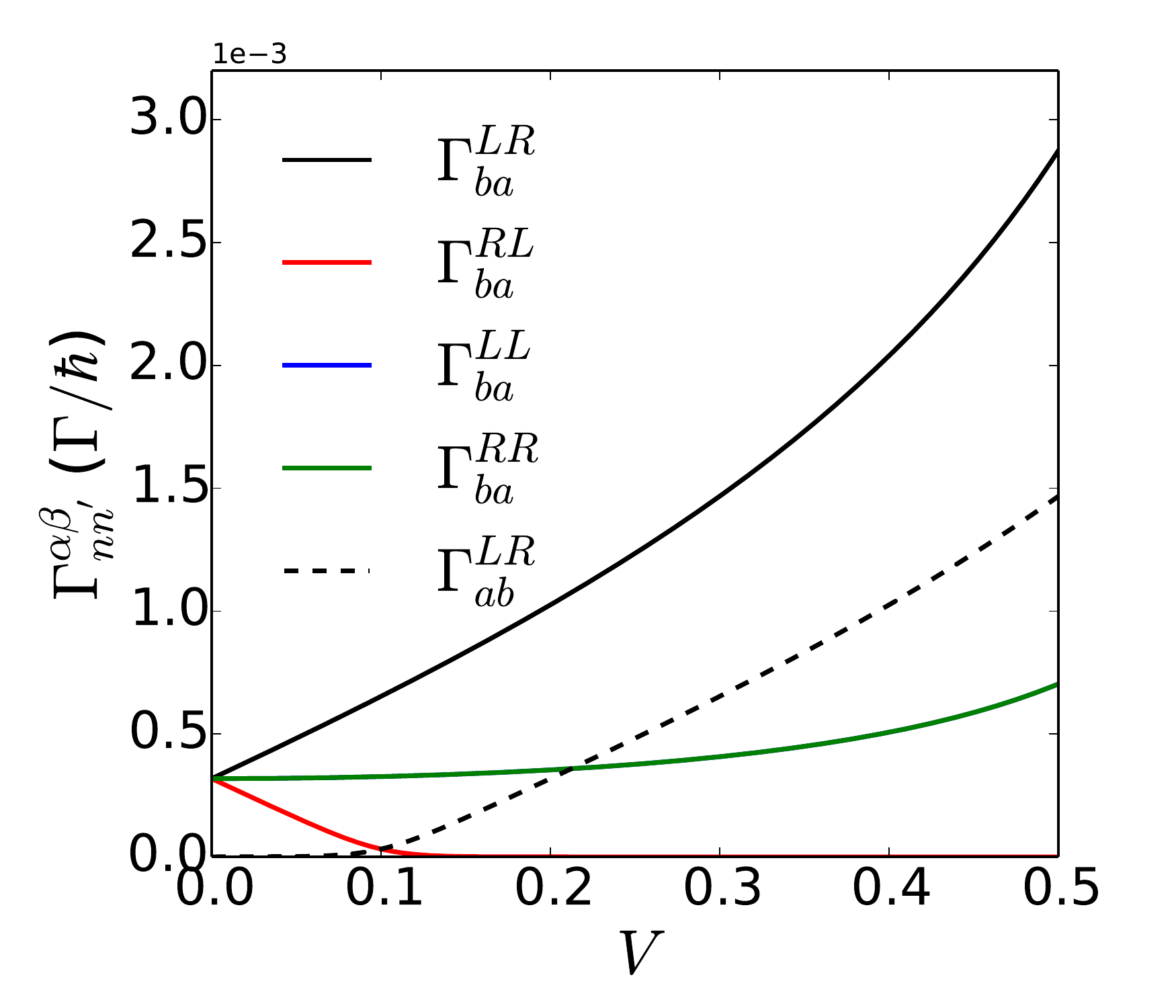}} \hspace{1cm}
  \subcaptionbox*{\bf(c)}[0.25\linewidth]{
    \includegraphics[width=0.25\linewidth]{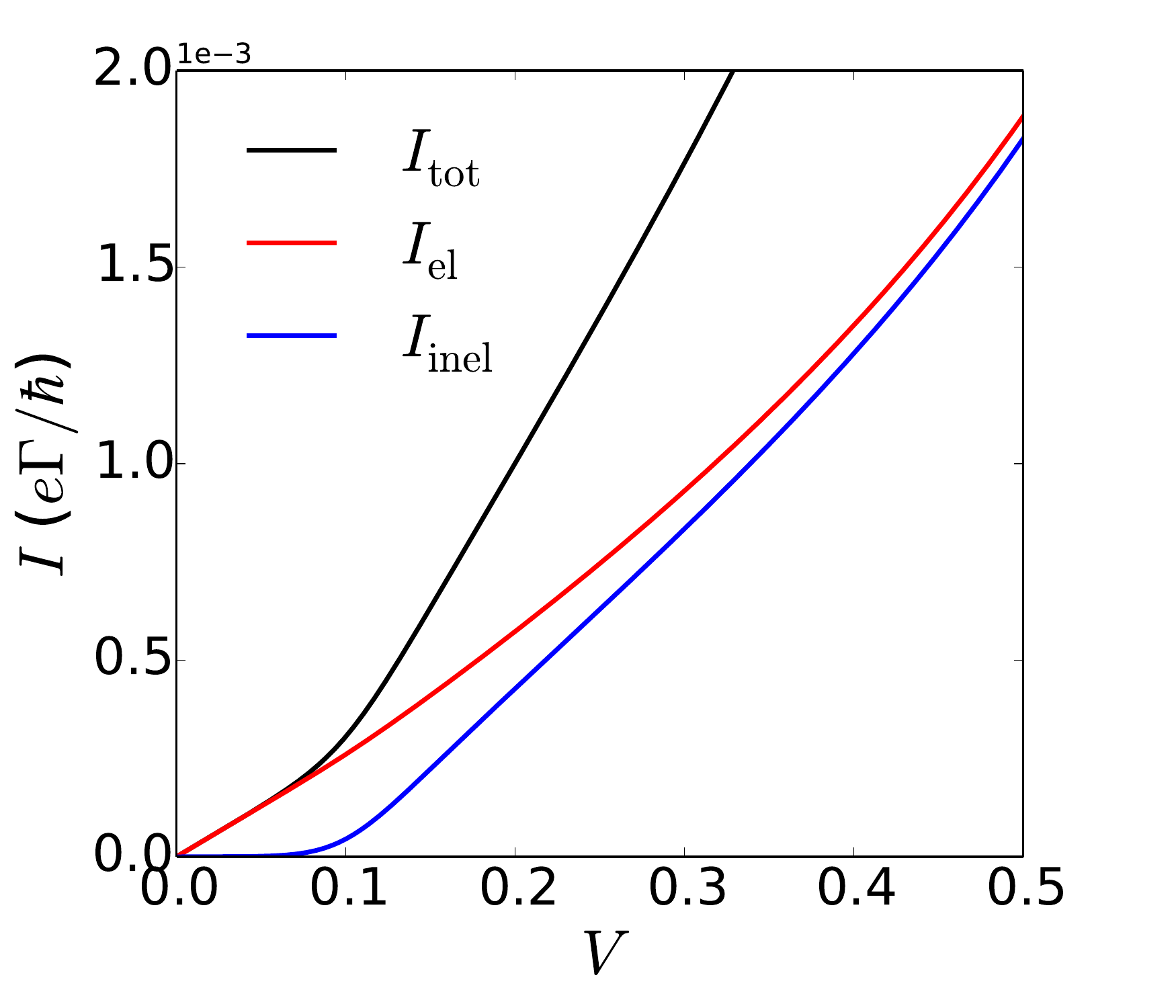}}     \hspace{1cm}
  \subcaptionbox*{\bf(e)}[0.25\linewidth]{
    \includegraphics[width=0.25\linewidth]{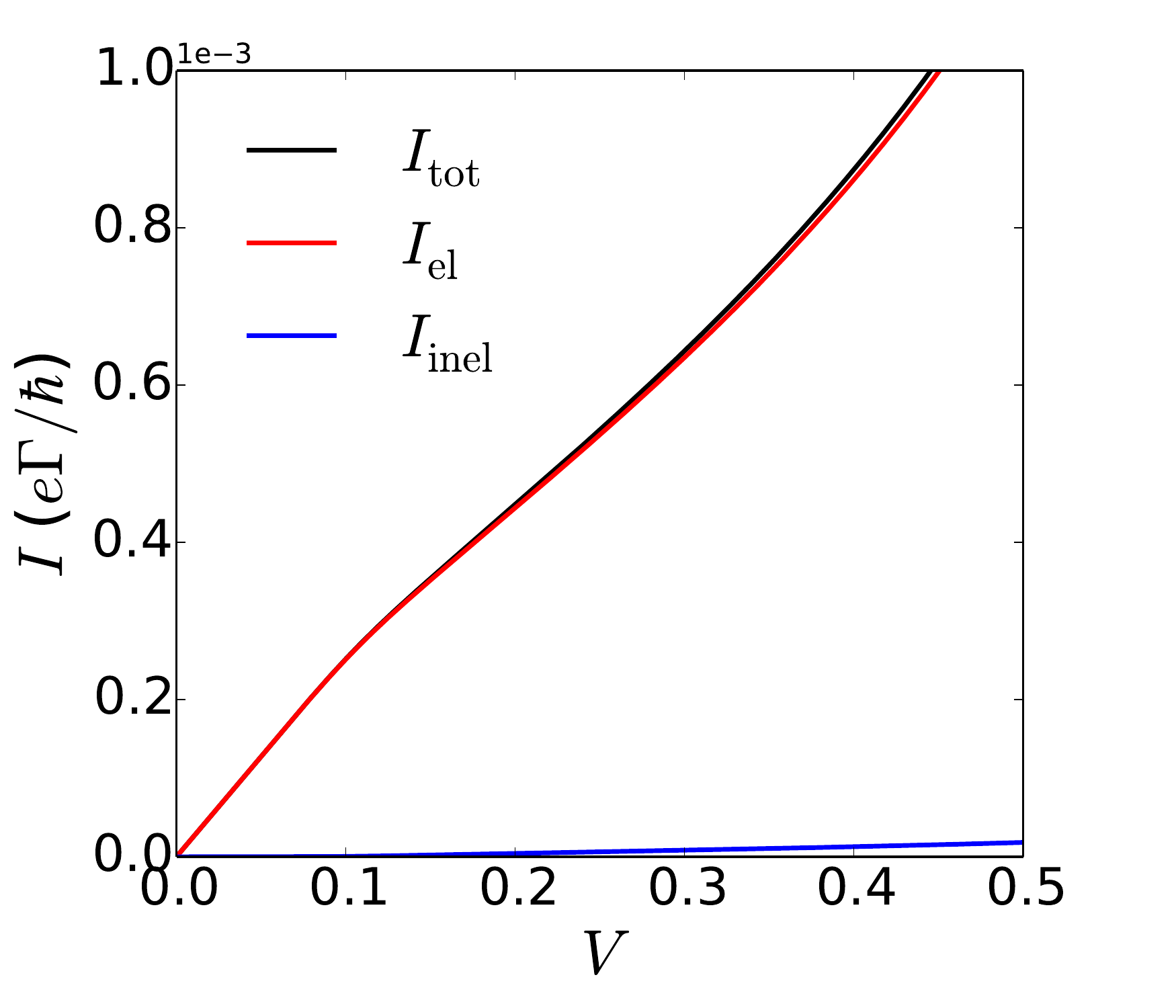}}
  \subcaptionbox*{\hspace{.3cm}\bf(b)}[0.25\linewidth]{
    \includegraphics[width=0.25\linewidth]{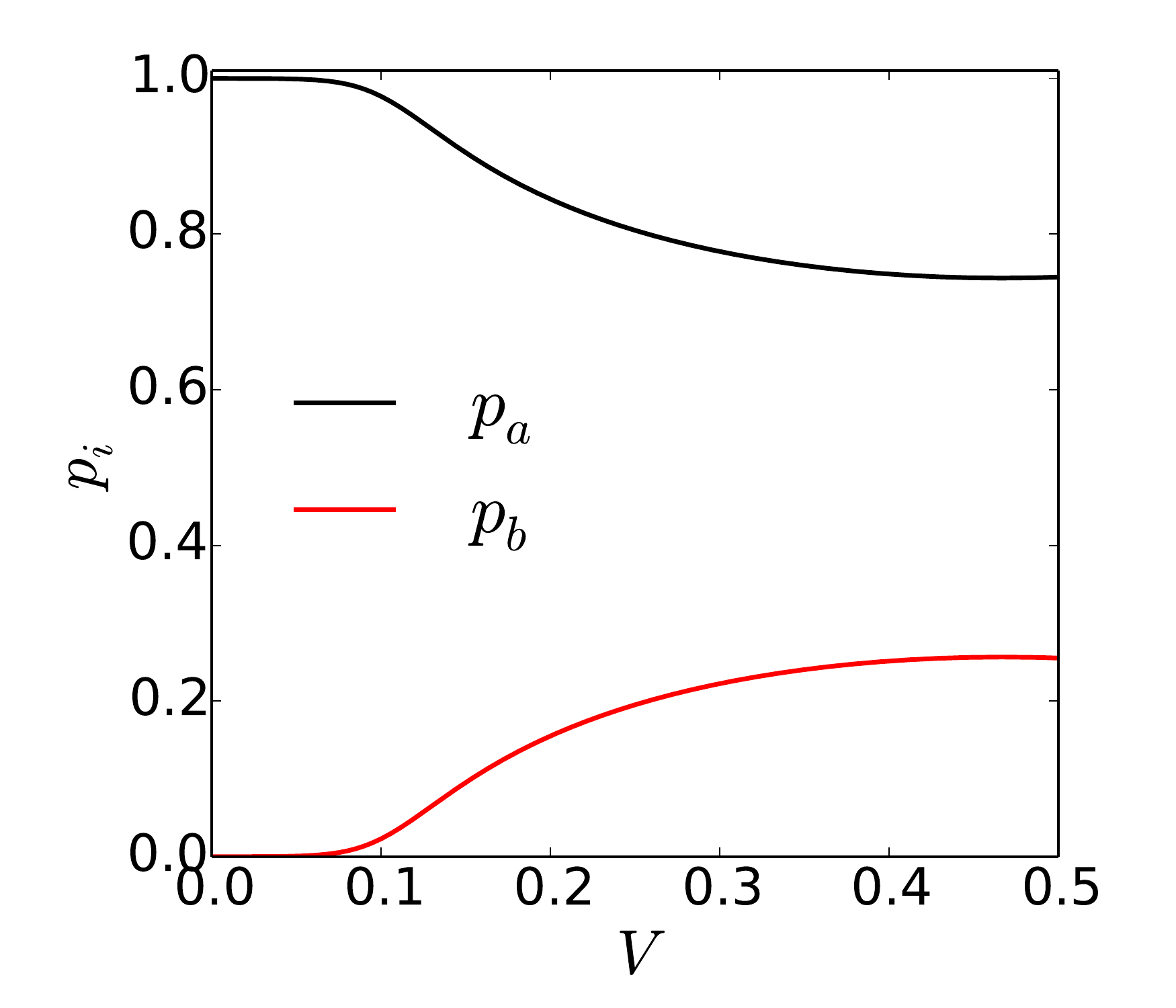}}    \hspace{1cm}
  \subcaptionbox*{\bf(d)}[0.25\linewidth]{
    \includegraphics[width=0.25\linewidth]{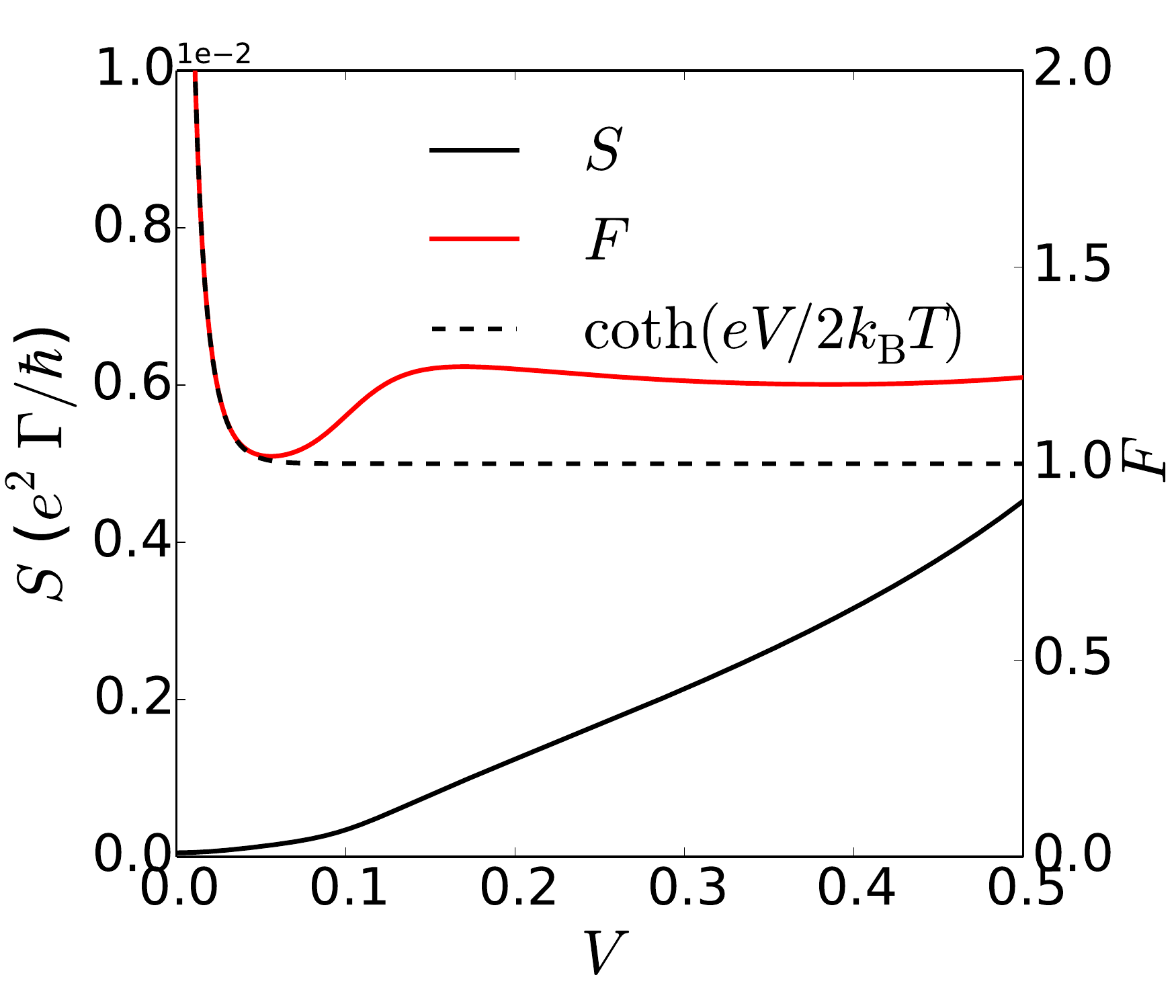}} \hspace{1cm}
  \subcaptionbox*{\bf(f)}[0.25\linewidth]{
    \includegraphics[width=0.25\linewidth]{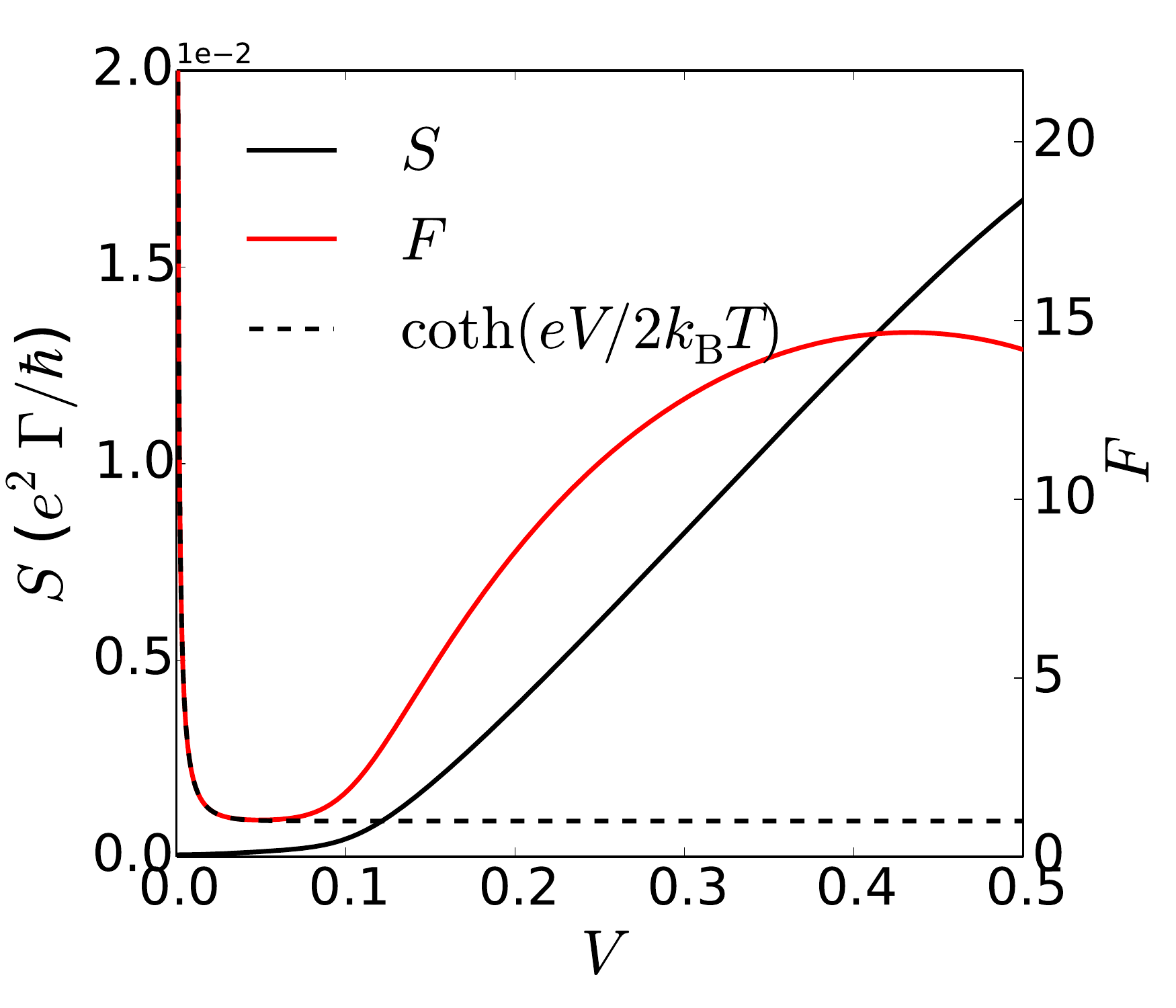}}
  \caption{(Color online) Super-Poissonian noise in the cotunneling regime. (a)
    Rates for inelastic cotunneling between the $\ket{N a}$, $\ket{N b}$ states
    as a function of bias. (b) Occupation probabilities for the $\ket{N a/b}$
    states. (c),(d) Cotunneling current [(c)], shot noise [(d) left axis] and
    Fano factor [(d) right axis] vs bias voltage. (e),(f) Same as in (c),(d) for
    the situation where the excited state is a blocking state. The plots
    correspond to the low-bias part of the cuts indicated by dashed lines in
    Fig.~\ref{fig:twolevel_stability}(b) [(a)--(d)] and
    Fig.~\ref{fig:twolevel_stability}(d) [(e)+(f)].}
\label{fig:twostate}
\end{figure*}

The corresponding stability diagrams for the Fano factor are shown in the lower
row of Fig.~\ref{fig:twolevel_stability}. Like the conductance, the Fano factors
are highly sensitive to the matrix elements of the blocking state. The Fano
factors have pronounced features with strong super-Poissonian values in the
blockaded regions of the stability diagrams where the signature in the current
and conductance is weak. These features originate from the opening of
cotunneling related transport channels. In particular, at the threshold for
inelastic cotunneling at $V=\Delta$, the Fano factor increases markedly. Also,
at the onset of COSET processes near the edges of the blockaded region, the Fano
factor shows drastic changes. In the regions outside the blockaded regions where
NDC occurs, strong super-Poissonian noise with a large Fano factor is observed.
The Fano factor in Fig.~\ref{fig:twolevel_stability}(f) where the excited state
is only partially blocked, shows a mixture of the features present in
Figs.~\ref{fig:twolevel_stability}(b) and~\ref{fig:twolevel_stability}(d)
for no and complete blocking, respectively.

In the following subsection, we analyze the shot noise in the different
transport regimes in closer detail.

\subsubsection{Super-Poissonian noise in the cotunneling regime} 

We start by considering the noise in the cotunneling regime (low-bias part of
the cuts through the center of the diamonds in
Fig.~\ref{fig:twolevel_stability}). In this regime, the results based on a pure
cotunneling description from Sec.~\ref{sec:twolevel} apply
[Eqs.~\eqref{eq:I_twolevel} and~\eqref{eq:S_twolevel}].

Figure~\ref{fig:twostate}(a)--\ref{fig:twostate}(d) summarize the situation
without a blocking state corresponding to the stability diagrams in
Figs.~\ref{fig:twolevel_stability}(a) and~\ref{fig:twolevel_stability}(b). The
inelastic cotunneling rates are shown in Fig.~\ref{fig:twostate}(a) as a
function of bias voltage for the different lead indices. At low temperature,
inelastic cotunneling processes in the direction of the voltage drop
($L\rightarrow R$) may excite the QD when $V>\Delta$. At the threshold
$V=\Delta$, the corresponding rate $\Gamma_{N,ab}^{LR}$ starts to increase
linearly with the applied bias. De-excitation processes with rates
$\Gamma_{N,ba}^{\alpha\beta}$, which relax the QD back to its ground state, are
always possible. The resulting occupation probabilities given by
Eq.~\eqref{eq:p_twolevel} are shown in Fig.~\ref{fig:twostate}(b) with the
probability for the ground (excited) state decreasing (increasing) approximately
linearly with the bias near the threshold $V\gtrsim \Delta$.

The occupation of the excited state at $V>\Delta$ gives rise to a strong
inelastic signal in the current [Fig.~\ref{fig:twostate}(c)] corresponding to a
positive step in the differential conductance $dI/dV$. At high bias voltage
$V\gg\Delta$, the elastic and inelastic contributions to the total current in
Eq.~\eqref{eq:I_twolevel} become equal. The shot noise shown in
Fig.~\ref{fig:twostate}(d) (left axis) together with the Fano factor (right
axis), also shows a clear signal at the inelastic threshold. The Fano factor is
given by its equilibrium value $F\sim \coth (eV / 2k_B T)$ at low bias $V
\lesssim k_B T$, drops to the Poissonian value $F=1$ for $k_B T < V < \Delta$,
and becomes super-Poissonian with $F>1$ for $V > \Delta$. The modest value of
the super-Poissonian Fano factor ($F \sim 1.2$) stems from the fact that the
elastic and inelastic cotunneling rates are of the same order of magnitude.

Figures~\ref{fig:twostate}(e) and~\ref{fig:twostate}(f) show the current and
shot noise in the case where $\ket{Nb}$ is a blocking state [cut along dashed
line in Fig.~\ref{fig:twolevel_stability}(d)]. In this case, the elastic and
inelastic cotunneling rates involving $\ket{Nb}$ are strongly
reduced. Therefore, the current in Fig.~\ref{fig:twostate}(e) is completely
dominated by the elastic component through $\ket{Na}$, and the corresponding
differential conductance has a negative step at the inelastic threshold where
the badly conducting excited state becomes populated. The noise in
Fig.~\ref{fig:twostate}(f), on the other hand, increases non-linearly with the
bias for $V > \Delta$. As a consequence, the Fano factor becomes strongly
super-Poissonian with $F\gg 1$. The mechanism behind the increased noise can be
understood as telegraphic switching between two elastic current channels with
different intrinsic shot noise.

We emphasize that the appearance of strong super-Poissonian noise in the
cotunneling regime relies on the excited $N$-electron state being blocked from
both the $N\pm1$ states, i.e. all the matrix elements $M_{Nb,N\pm1}^{L/R}$ must
be small. In addition, the relaxation rate due to coupling to an external
equilibrium bath in Eq.~\eqref{eq:Gamma_relax} must be small compared to the
inelastic cotunneling rates, $\Gamma_{ab/ba}^\text{rel} \ll
\Gamma_{N,ab/ba}^{\alpha\beta}$. If this is not the case, the QD relaxes to its
ground state on a time scale much faster than the time between cotunneling
events. Hence, elastic cotunneling via the ground state will dominate the
current and noise and the Fano factor becomes Poissonian $F\sim 1$ in the limit
of strong external relaxation.

\subsubsection{Noise in the COSET regime}

\begin{figure}[!t]
  \captionsetup[sub]{justification=centering,position=top,font=small,margin=0pt,skip=-2pt}
  \subcaptionbox*{\hspace{.25cm}\bf(a)}[0.49\linewidth]{
    \includegraphics[width=0.49\linewidth]{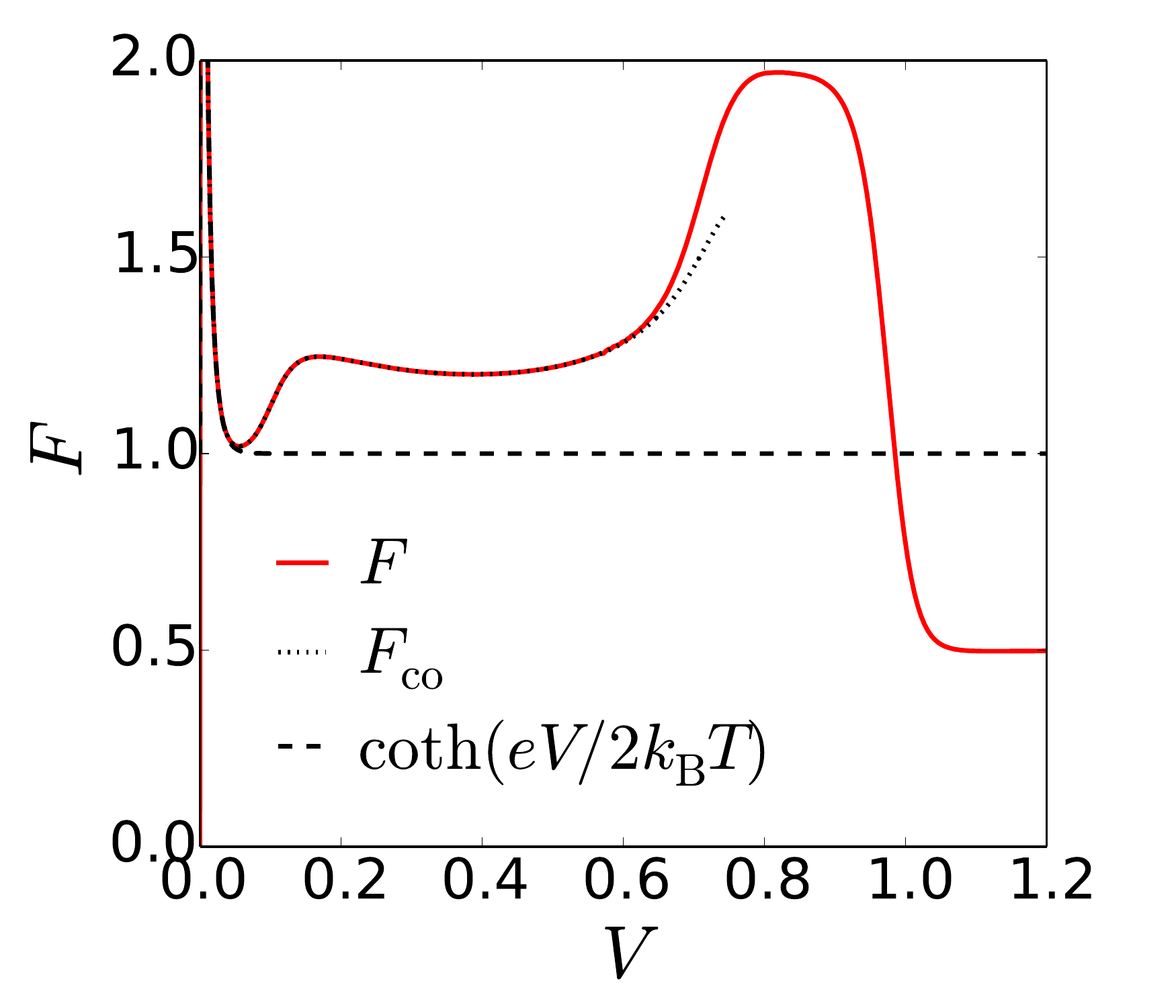}}
  \subcaptionbox*{\hspace{.25cm}\bf(b)}[0.49\linewidth]{
    \includegraphics[width=0.49\linewidth]{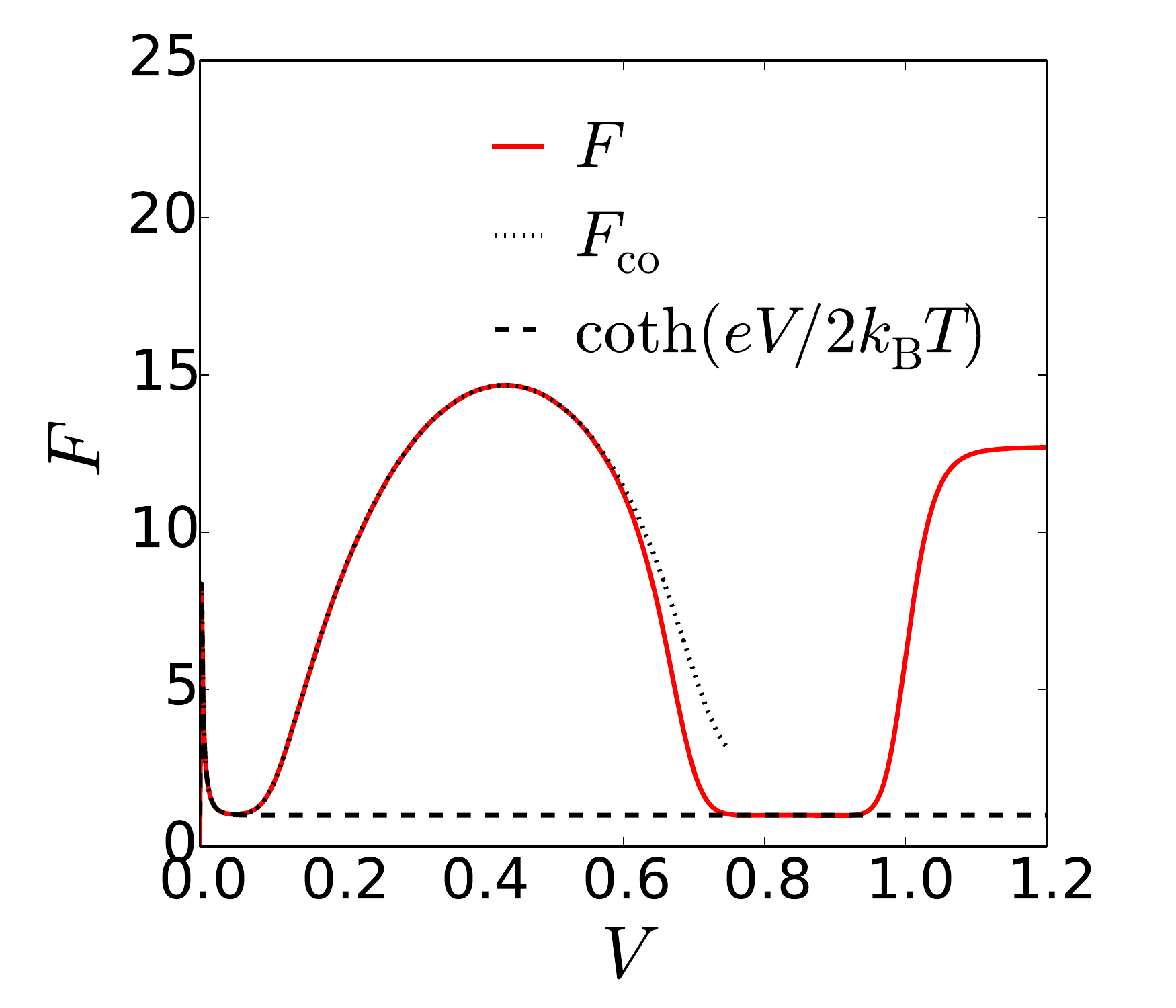}}
    \caption{(Color online) Fano factor in the cotunneling and COSET regime
      along the cuts marked with dashed lines in the stability diagrams of
      Fig.~\ref{fig:twolevel_stability}. The Fano factor $F_\text{co}$ obtained
      from the pure cotunneling expressions for the current and noise in
      Eqs.~\eqref{eq:I_twolevel} and~\eqref{eq:S_twolevel} is also shown. Above
      the onset of COSET processes a pure cotunneling description becomes
      ill defined. The plots correspond to the cuts indicated by dashed lines in
      Fig.~\ref{fig:twolevel_stability}(b) [(a)] and
      Fig.~\ref{fig:twolevel_stability}(d) [(b)].}
\label{fig:twolevel_cotunneling_cuts}
\end{figure}

To get a clearer picture of the behavior of the noise in the part of the Coulomb
blockaded region where cotunneling and sequential tunneling coexist and COSET
processes provide a relaxation channel for the excited state, we show in
Fig.~\ref{fig:twolevel_cotunneling_cuts} the bias dependence of the Fano factor
along the full cuts (positive bias only) marked with dashed lines in
Fig.~\ref{fig:twolevel_stability}. For comparison, the dotted lines in
Fig.~\ref{fig:twolevel_cotunneling_cuts} show the Fano factor $F_\text{co}$
obtained from Eqs.~\eqref{eq:I_twolevel} and~\eqref{eq:S_twolevel} taking into
account cotunneling only. At low bias, the noise is dominated by cotunneling
processes. As discussed above, the noise becomes super-Poissonian at the
threshold for inelastic cotunneling at $V=\Delta$ and acquires a strongly
super-Poissonian Fano factor $F\sim 15$ in the presence of the blocking state.

The onset of COSET processes takes place at the side-band resonances at
$V = 2 \abs{\tilde{\varepsilon}_0 + U - \Delta}$ and
$V = 2\abs{\tilde{\varepsilon}_0 + \Delta}$, where relaxation of the excited
state $\ket{Nb}$ via sequential tunneling to the $\ket{N\pm 1}$ states becomes
possible. In the COSET regime, the change in the Fano factor is qualitative in
the two cases. Whereas it increases to a value of $F\sim 2$ without, it drops to
$F=1$ with a blocking state. The increase in the Fano factor in the former case,
is due to COSET processes where a channel for sequential tunneling inside the
blockaded region opens when the excited state becomes populated via inelastic
cotunneling~\cite{Schon:CAST}. This gives rise to a current that is alternately
governed by sequential and cotunneling every time an inelastic cotunneling
process excites and de-excites the QD, respectively. However, in the presence of
the blocking state, sequential and cotunneling via the excited state are
strongly suppressed. The current is therefore dominated by elastic cotunneling
via the ground state and the shot noise becomes Poissonian with $F=1$. This is
markedly different from the situation without a blocking
state~\cite{Schon:CAST}. At bias voltages $V> 2\abs{\tilde{\varepsilon}_0}$ and
$V>2\abs{\tilde{\varepsilon}_0 + U}$, the main resonances enter the bias window
and sequential tunneling becomes dominant.

\subsubsection{Sub-Poissonian noise, NDC and super-Poissonian telegraphic noise
  in the resonant regime}

\begin{figure}[!t]
  \captionsetup[sub]{justification=centering,position=top,font=small,margin=0pt,skip=-2pt}
  \subcaptionbox*{\hspace{.25cm}\bf(a)}[0.49\linewidth]{
    \includegraphics[width=0.49\linewidth]{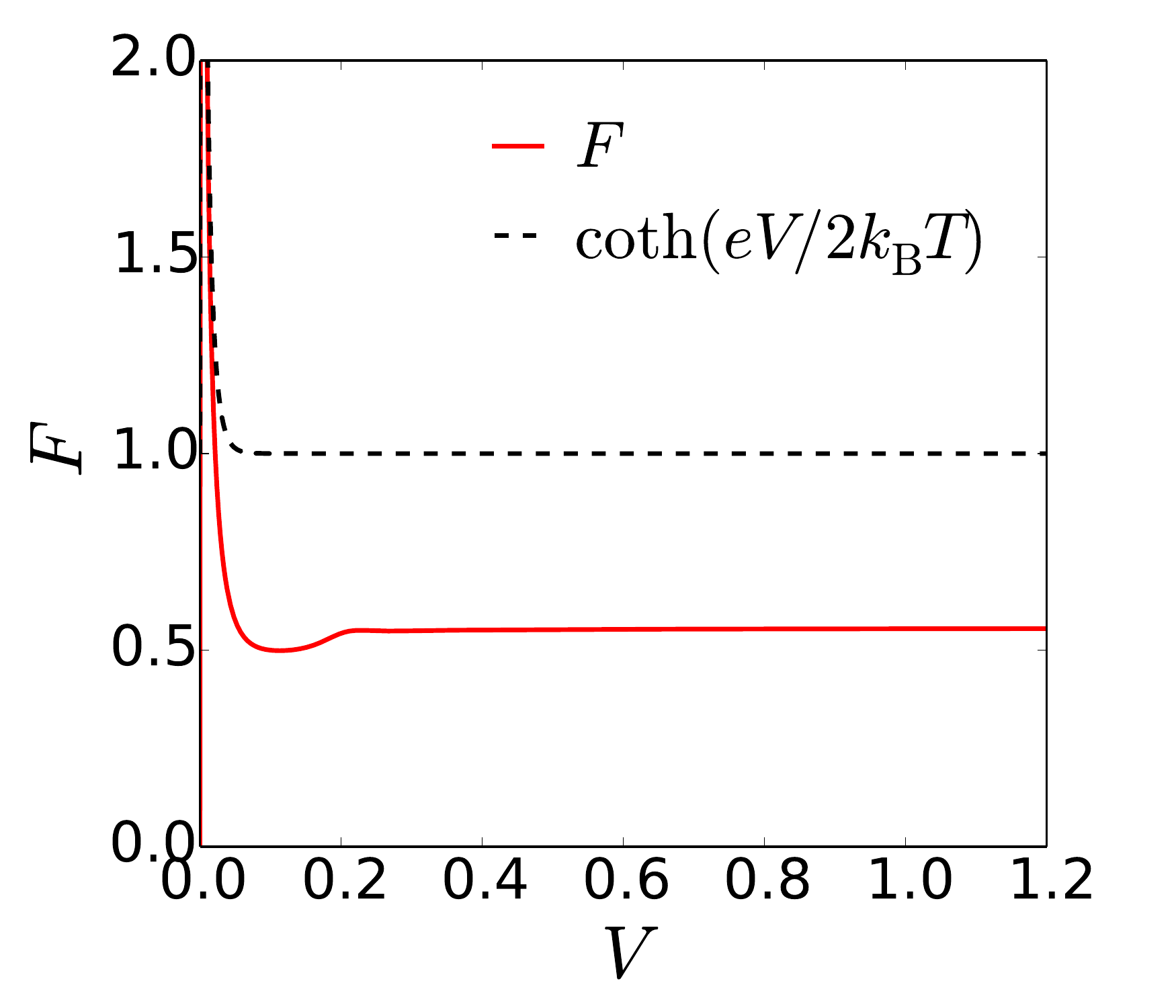}}
  \subcaptionbox*{\hspace{.25cm}\bf(b)}[0.49\linewidth]{
    \includegraphics[width=0.49\linewidth]{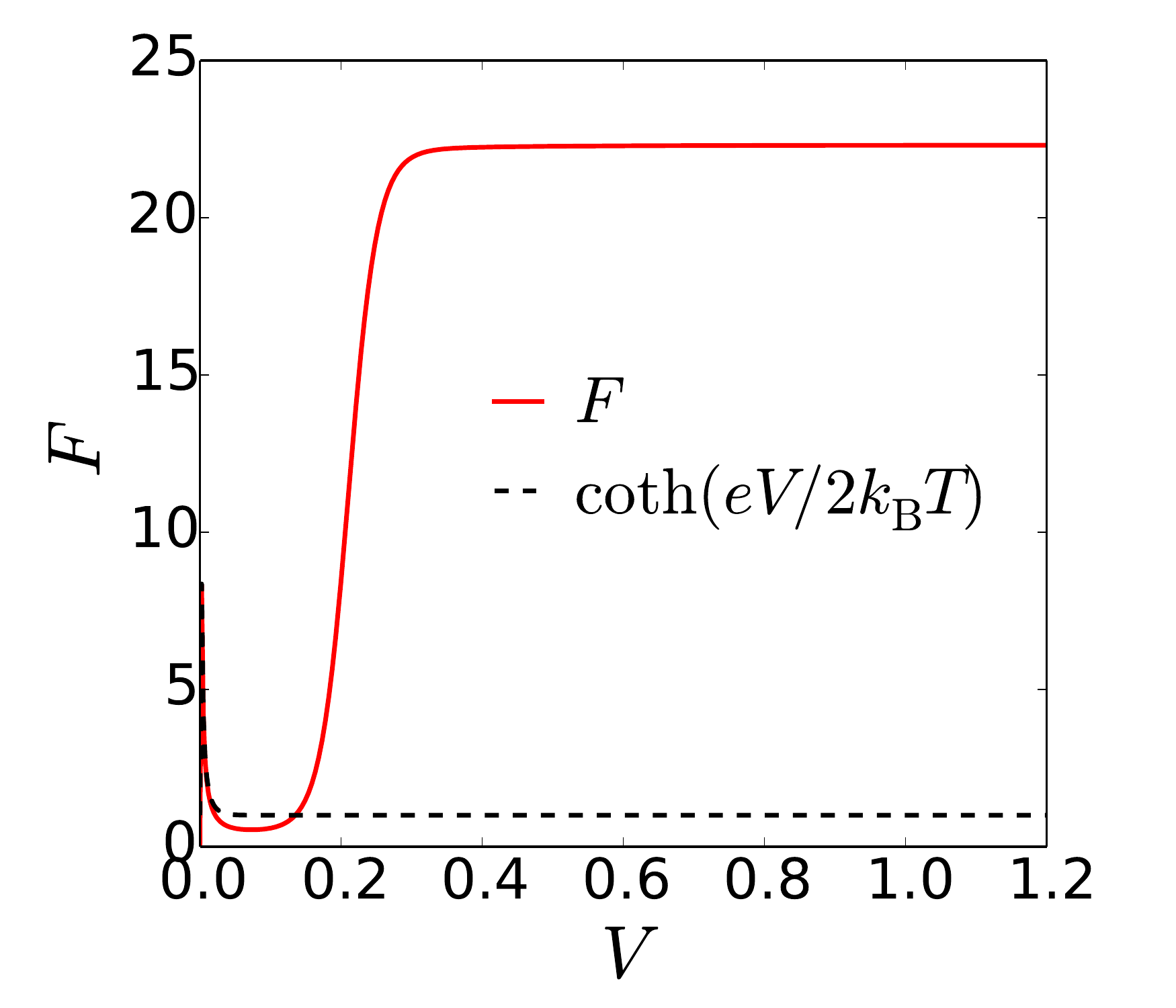}}
    \caption{(Color online) Fano factor in the resonant regime along the cuts
      marked with dotted lines in the stability diagrams of
      Fig.~\ref{fig:twolevel_stability}. The plots correspond to the cuts
      indicated by dotted lines in Fig.~\ref{fig:twolevel_stability}(b) [(a)]
      and Fig.~\ref{fig:twolevel_stability}(d) [(b)].}
\label{fig:twolevel_ndr_cuts}
\end{figure}
Figure~\ref{fig:twolevel_ndr_cuts} shows the bias dependence of the Fano factor
along the cuts in the resonant regime outside the Coulomb blockaded regions
marked with dashed lines in Fig.~\ref{fig:twolevel_stability}. In the situation
without a blocking state, the Fano factor is sub-Poissonian with $F=0.5$ for
$k_B T < V < 2 \Delta$ and a slightly larger value $F\sim0.55$ for $V > 2
\Delta$, which is characteristic for sequential tunneling through a QD with
excited states~\cite{Nazarov:FCS,Schon:ShotNoise,Belzig:Full}. However, in the
presence of the blocking state, the shot noise becomes strongly super-Poissonian
with $F\sim 23$. The mechanism responsible for NDC and the strong enhancement
of the noise is the same. When the blocking state enters the bias window, the
QD gets trapped in the blocking state due to the small transition rate to other
states. This reduces the current and results in a telegraphic noise with long
quiet periods without charge transfer interrupted by avalanches of transfer
processes every time the QD escapes the blocking state.

\section{Summary}

In summary, we have demonstrated how the standard scheme to evaluate the FCS of
charge transfer due to sequential tunneling in Coulomb blockaded QD
systems~\cite{Nazarov:FCS} can be generalized to take into account cotunneling
processes. In analogy with the procedure for sequential tunneling, this is done
by replacing the cotunneling rates in the Markovian master
equation~\eqref{eq:rateequation} with counting-field dependent rates as
described in Eqs.~\eqref{eq:dotp_chi} and~\eqref{eq:Gamma_chi}. This approach
neglects non-Markovian
effects~\cite{Schon:Cotunneling,Fazio:NonMarkovian,Schon:Intermediate}
associated with tunneling-induced level broadening and shifts, and, hence, only
applies for $k_B T,eV\gg \Gamma$ or in the cotunneling regime, $\delta \gg
\Gamma$. In the cotunneling regime, we have demonstrated that the results for
shot noise and the FCS from more elaborate
methods~\cite{Schon:Cotunneling,Fazio:NonMarkovian} are reproduced. In addition,
we have obtained an analytic expression for the CGF
[Eq.~\eqref{eq:CGF_twolevel}] describing the charge-transfer statistics of
elastic and inelastic cotunneling in a two-state QD system.

Studying a generic model for a QD with an excited electronic state, we found
that the shot noise in the cotunneling regime is inherently super-Poissonian for
voltages larger than the inelastic threshold $V>\Delta$. A strongly enhanced
noise level with Fano factor $F\gg 1$ results when the excited state is a
so-called blocking state. This is due to telegraphic switching between the two
differently conducting channels for elastic cotunneling via the ground and
excited state. In the presence of environmental relaxation, the super-Poissonian
noise is reduced and becomes Poissonian with $F=1$ once the relaxation dominates
the inelastic cotunneling rates. In the COSET regime where cotunneling and
sequential tunneling coexist, we found that the noise is, respectively,
super-Poissonian and Poissonian for an excited state without and with blocking
properties.

Our approach for evaluating the FCS can be generalized to other higher-order
tunneling processes in QD systems, such as, e.g., pair
tunneling~\cite{Oppen:Pair,Hettler:Pair} and charge reconfiguration processes in
multi-dot
systems~\cite{Hawrylak:Stability,Hawrylak:Physics,Mahalu:Frustration,Sim:Anisotropic},
and may be applied to investigate, e.g., the interplay between inelastic
cotunneling and quantum interference in the FCS of molecular
contacts~\cite{Paaske:Quantum}.

\begin{acknowledgments}
  We would like to thank A.~Nitzan, T.~Novotn{\'y}, and C.~Flindt for fruitful
  discussions. W.~B. acknowledges financial support by the DFG through SFB 767,
  the Kurt Lion Foundation, and an EDEN (Erasmus Mundus Academic Network) grant.
  K.K. acknowledges support from the Villum and Carlsberg Foundations.
\end{acknowledgments}

\vspace{1cm}

\appendix

\section{Master equation for elastic and inelastic cotunneling in a two-state system}
\label{sec:twolevel_rateequation}

In this case, the standard master equations takes the form of the $2\times 2$ matrix
\begin{equation}
  \label{eq:M_twolevel}
  \mathbf{M} = 
   \begin{pmatrix}
     - \Gamma_{01} & \Gamma_{10}  \\
     \Gamma_{01}   & -\Gamma_{10}  
  \end{pmatrix} ,
\end{equation}
where $\Gamma_{ij} = \sum_{\alpha\beta} \Gamma_{ij}^{\alpha\beta}$ are the rates
for the inelastic cotunneling-induced transitions between the states.

The eigenvalues are found to be
\begin{equation}
  \lambda = \frac{1}{2} 
      \left[ -\left( \Gamma_{10} + \Gamma_{01} \right) \pm 
        \left( \Gamma_{10} + \Gamma_{01} \right)
      \right]  ,
\end{equation}
with the eigenvector of the zero eigenvalue giving the steady-state solution,
\begin{equation}
  \mathbf{p}_{\lambda=0} =
  \left( 
    \frac{\Gamma_{10}}{\Gamma_{01} + \Gamma_{10}} ,
    \frac{\Gamma_{01}}{\Gamma_{01} + \Gamma_{10}} 
  \right)^T .
\end{equation}
The current can be obtained from the steady-state solution as
\begin{align}
  \label{eq:current_twolevel}
  I & = p_0 \left( \Gamma_{00}^{LR} - \Gamma_{00}^{RL} \right) + 
        p_1 \left( \Gamma_{11}^{LR} - \Gamma_{11}^{RL} \right) 
        \nonumber \\
    & \quad +
        p_0 \left( \Gamma^{LR}_{01} - \Gamma^{RL}_{01} \right) + 
        p_1 \left( \Gamma^{LR}_{10} - \Gamma^{RL}_{10} \right) ,
        \nonumber \\
    & = I_\text{el} + I_\text{inel} ,
\end{align}
where the first and second line are the elastic $I_\text{el}$ and inelastic
$I_\text{inel}$ contributions to the current, respectively.

\begin{widetext}
\section{Regularized cotunneling rates}
\label{sec:regularization}

The integrals for the elastic and inelastic cotunneling rates can be evaluated
using the standard regularization scheme. The procedure for regularizing the
diverging integrands in the cotunneling rates of
Eqs.~\eqref{eq:Gamma_elastic1}--\eqref{eq:Gamma_inelastic2} can be found in
Ref.~\onlinecite{Andreev:FranckCondon}. 

The integrants in the expressions for the cotunneling rates can be written on
the general form
\begin{equation}
  \left\vert
    \frac{A}{\varepsilon - \varepsilon_1 + i0^+} \pm
    \frac{B}{\varepsilon - \varepsilon_2 + i0^+}
  \right\vert^2 = 
  \left\vert
    \frac{A}{\varepsilon - \varepsilon_1 + i0^+}
  \right\vert^2 +
  \left\vert
    \frac{B}{\varepsilon - \varepsilon_2 + i0^+}
  \right\vert^2 \pm
  2 \text{Re}\left(
    \frac{A}{\varepsilon - \varepsilon_1 + i0^+} 
    \frac{B}{\varepsilon - \varepsilon_2 - i0^+}
  \right) ,
\end{equation}
where we have added a infinitesimal broadening of the QD states (the
regularizer) in the denominators.

The regularized rates can then be obtained analytically from the following two
integrals,
\begin{align}
  \label{eq:integral1}
  \int   d\varepsilon \, \frac{f(\varepsilon - E_1) [1- f(\varepsilon-E_2)]}
     {(\varepsilon - \varepsilon_1) ( \varepsilon - \varepsilon_2)} 
     = \frac{n_B(E_2 - E_1)}{\varepsilon_1 - \varepsilon_2} 
       \,\mathrm{Re}\left[ \psi(E_{21}^+) - \psi(E_{22}^-) - \psi(E_{11}^+)
                         + \psi(E_{12}^-)  \right] 
\end{align}
and 
\begin{align}
  \label{eq:integral2}
  \int d\varepsilon \, \frac{f(\varepsilon - E_1) [1- f(\varepsilon-E_2)]}
      {(\varepsilon - \varepsilon_1)^2} 
      = \frac{n_B(E_2 - E_1)}{2\pi k_BT}
        \, \mathrm{Im}\left[ \psi'(E_{21}^+) - \psi'(E_{11}^+) \right] ,
\end{align}
where $\psi$ denotes the digamma function, $\psi'$ its derivative, $E_{ij}^\pm =
\tfrac{1}{2} \pm \tfrac{i}{2\pi k_BT}(E_i - \varepsilon_j)$, and
$n_B$ the Bose-Einstein distribution.
\end{widetext}

\bibliography{paper.bbl}

\end{document}